\documentclass[12pt]{article}

\usepackage{geometry} 

\usepackage{graphicx}
\usepackage{tabularx}
\usepackage{typearea}
\usepackage{wrapfig}
\usepackage{fancyhdr}
\usepackage{amsmath}
\usepackage{amssymb}
\usepackage{amsthm}

\usepackage{accents}
\usepackage[ansinew]{inputenc}
\usepackage[arrow, matrix, curve]{xy}
\usepackage{enumerate}
\usepackage{bbm}
\usepackage{mathtools}
\usepackage{cite}
\usepackage{tikz}
\usepackage{etex}
\usepackage{tikz-qtree}
\usetikzlibrary{trees}
\usetikzlibrary{calc}

\numberwithin{equation}{section}

\theoremstyle{theorem}
\newtheorem{Theorem}{Theorem}[section]
\newtheorem{Proposition}[Theorem]{Proposition}
\newtheorem{Lemma}[Theorem]{Lemma}
\newtheorem{Corollary}[Theorem]{Corollary}
\newtheorem{Definition}[Theorem]{Definition}

\theoremstyle{remark}
\newtheorem{Remark}[Theorem]{Remark}
\newtheorem{Example}[Theorem]{Example}
\newcommand{\bco}{\begin{Corollary}}
\newcommand{\eco}{\end{Corollary}}

\newcommand{\BF}[3]{(\chi_{#1})^{#2}_{#3}}
\newcommand{\BFC}[3]{(\chi_{#1}^{\CC})^{#2}_{#3}}
\newcommand{\BFi}[1]{\chi_{#1}}
\newcommand{\BFCi}[1]{\chi_{#1}^{\CC}}
\newcommand{\bpr}{\begin{Proposition}}
\newcommand{\epr}{\end{Proposition}}

\newcommand{\btm}{\begin{Theorem}}
\newcommand{\etm}{\end{Theorem}}

\newcommand{\ben}{\begin{enumerate}}
\newcommand{\een}{\end{enumerate}}

\newcommand{\bit}{\begin{itemize}}
\newcommand{\eit}{\end{itemize}}

\newcommand{\bca}{\begin{cases}}
\newcommand{\eca}{\end{cases}}

\newcommand{\bre}{\begin{Remark}\rm}
\newcommand{\ere}{\end{Remark}}

\newcommand*{\bbm}{\begin{Remark}}
\newcommand*{\ebm}{\end{Remark}}

\newcommand{\ble}{\begin{Lemma}}
\newcommand{\ele}{\end{Lemma}}

\newcommand*{\bsz}{\begin{Proposition}}
\newcommand*{\esz}{\end{Proposition}}

\newcommand{\beq}{\begin{equation}}
\newcommand{\eeq}{\end{equation}}

\newcommand{\bbma}{\begin{bmatrix}}
\newcommand{\ebma}{\end{bmatrix}}

\newcommand{\bpma}{\begin{pmatrix}}
\newcommand{\epma}{\end{pmatrix}}

\newcommand*{\bbs}{\begin{Example}}
\newcommand*{\ebs}{\end{Example}}

\newcommand*{\bfg}{\begin{Corollary}}
\newcommand*{\efg}{\end{Corollary}}

\newcommand*{\bdf}{\begin{Definition}}
\newcommand*{\edf}{\end{Definition}}

\newcommand*{\bbw}{\begin{proof}}
\newcommand*{\ebw}{\end{proof}}

\newcommand*{\bpf}{\begin{proof}}
\newcommand*{\epf}{\end{proof}}

\newtagform{simple}{}{}
\newtagform{standard}{(}{)}
\newcommand{\eqqed}{\usetagform{simple}\tag{$\square$}}
\newcommand{\eqqedan}{\usetagform{simple}}
\newcommand{\eqqedaus}{\usetagform{standard}}

\allowdisplaybreaks

\newcommand{\II}{\mathbbm{1}}

\newcommand{\CC}{{\mathbb{C}}}

\newcommand{\PP}{{\mathbb{P}}}

\newcommand{\momap}{\mu}

\newcommand{\pr}{\mr{pr}}

\newcommand*{\res}{\upharpoonright}

\newcommand{\tinyfrac}[2]{\text{\tiny $\frac{#1}{#2}$}}
\newcommand{\sitem}{\rm\item\it}

\newcommand{\SL}{{\mr{SL}}}
\renewcommand{\sl}{\mf{sl}}

\newcommand{\SU}{{\mr{SU}}}
\newcommand{\su}{\mf{su}}

\newcommand{\al}[1]{\begin{align} #1 \end{align}}
\newcommand{\ala}[1]{\begin{align*} #1 \end{align*}}

\DeclareMathOperator{\Ad}{Ad}

\DeclareMathOperator{\id}{id}
\DeclareMathOperator{\im}{im}

\DeclareMathOperator{\tr}{tr}

\newcommand{\mc}[1]{\mathcal{#1}}
\newcommand{\mf}[1]{\mathfrak{#1}}
\newcommand{\mr}[1]{\mathrm{#1}}

\newcommand{\comment}[1]{}
\newcommand{\verweis}[1]{}
\newcommand{\todo}[1]{}

\newcommand{\group}{G}

\newcommand{\pha}{{\mc P}}

\newcommand{\ket}[1]{|#1\rangle}
\newcommand{\bra}[1]{\langle#1|}
\newcommand{\braket}[2]{\langle#1|#2\rangle}
\newcommand{\ve}{\varepsilon}

\newcommand{\vp}{\varphi}

\newcommand{\ctg}{\mr T^\ast}

\newcommand{\rref}[1]{{\rm \ref{#1}}}

\newcommand{\ol}[1]{\overline{#1}}
\newcommand{\ul}[1]{\underline{#1}}
\newcommand{\ut}[1]{\undertilde{#1}}

\newcommand*{\qeb}{\nopagebreak\hspace*{0.1em}\hspace*{\fill}{\mbox{\small$\blacklozenge$}}}

\DeclareMathOperator{\ad}{ad}

\DeclareMathOperator{\End}{End}

\makeatletter
\newcommand{\douwidehat}[2]{%
  \sbox0{$\m@th#1\widehat{\hphantom{#2}}$}%
  \sbox2{$\m@th#1x$}
  \sbox4{$\m@th#1#2$}
  \dimen0=\ht0
  \advance\dimen0 -.8\ht2
  \dimen2=\dp4
  \rlap{%
    \raisebox{\dimexpr\dimen0-\dimen2}{%
      \scalebox{1}[-1]{\box0}%
    }%
  }%
  {#2}%
}
\makeatother  

\begin{document}

\title{The Hilbert space costratification for the orbit type strata of $\SU(2)$-lattice gauge theory}

 \author{
E. \ Fuchs$^\ast$, P.\ D.\ Jarvis$^\dagger$, G.\ Rudolph$^\ast$, M.\ Schmidt$^\ast$
\\[5pt]
$^\ast$ Institute for Theoretical Physics, University of Leipzig, 
\\
P.O. Box 100 920, D-4109 Leipzig, Germany.
\\[5pt]
$^\dagger$ School of Natural Sciences (Mathematics and Physics),\\  University of Tasmania,\\ Private Bag 37, GPO, Hobart Tas 7001, Australia. 
}

\date{\today}

\maketitle

\begin{abstract}

\noindent

We  construct the Hilbert space costratification of $G=\SU(2)$-quantum gauge theory on a finite spatial lattice in the Hamiltonian approach. We build on previous work \cite{FuRS}, 
where we have implemented the classical gauge orbit strata on quantum level within a suitable holomorphic picture. 
In this picture, each element $\tau$ of the classical stratification corresponds to the zero locus of a finite subset $\{p_i\}$ of the algebra $\mc R$ of $G$-invariant representative functions on 
$G^N_\CC$. Viewing the invariants as multiplication operators $\hat p_i$ on the Hilbert space $\mc H$, 
the union of their images defines a subspace  of $\mc H$ whose orthogonal complement $\mc H_\tau$ is the element of the costratification corresponding to $\tau$. 
To construct $\mc H_\tau$, one has to determine the images of the $\hat p_i$ explicitly. To accomplish that goal, we construct an orthonormal basis in $\mc H$ and determine 
the multiplication law for the basis elements, that is, we determine  the structure constants of $\mc R$ in this basis. This part of our analysis applies to any compact Lie group $G$. 
For $G = \SU(2)$, the above procedure boils down to a problem  
in combinatorics of angular momentum theory. Using this theory, we obtain the union of the images of the operators  $\hat p_i$ as a subspace generated by vectors whose 
coefficients with respect to our basis are given in terms of Wigner's $3nj$ symbols. The latter are further expressed in terms of $9j$ symbols. Using these techniques, we are also able to reduce the eigenvalue problem for the Hamiltonian of this theory to a problem in linear algebra. 

\end{abstract}

\newpage

\tableofcontents


\section{Introduction}


This paper is a continuation of our previous one \cite{FuRS}, where we have derived the defining relations for the orbit type strata of $G = \SU(2)$-lattice gauge theory. 
It is part of a program which aims at developing a non-perturbative approach to the quantum theory of gauge fields in the Hamiltonian framework with special emphasis on the role of non-generic gauge orbit types. The starting point is a finite-dimensional Hamiltonian lattice approximation of the theory, which on the classical level leads to a finite-dimensional Hamiltonian system with symmetries. The corresponding quantum theory is obtained via canonical quantization. It is best described in the language of $C^*$-algebras with a field algebra which (for a pure gauge theory) 
may be identified with the algebra of compact operators on the Hilbert space of square-integrable functions over the product $G^N$ of a number of copies of the gauge group manifold $G$.\footnote{We will see that $N$ is the number of off-tree links for a chosen maximal lattice tree.} Correspondingly, the observable algebra is obtained via gauge symmetry reduction. We refer to \cite{qcd1, qcd2, qcd3, RS} for the study of this algebra, including its superselection structure. 
For first steps towards the construction of the thermodynamical limit, see \cite{GR,GR2}. 

If the gauge group is non-Abelian, then the action of the symmetry group in the corresponding classical Hamiltonian system necessarily has more than one orbit type. Correspondingly, the reduced phase space obtained by symplectic reduction is a stratified symplectic space \cite{SjamaarLerman,OrtegaRatiu,RS} rather than a symplectic manifold as in the case with one orbit type \cite{AbrahamMarsden}. The stratification is given by the orbit type strata. It consists of an open and dense principal stratum and several secondary strata. Each of these strata is invariant under the dynamics with respect to any invariant Hamiltonian. For case studies we refer to  \cite{cfg, cfgtop,FRS}. 

To study the influence of the classical orbit type stratification on quantum level, we  use the concept of costratification of the quantum Hilbert space as developed by Huebsch\-mann 
\cite{Hue:Quantization}. A costratification is given by a family of closed subspaces, one for each stratum. Loosely speaking, the closed subspace associated with a certain classical stratum consists of the wave functions which are optimally localized at that stratum in the sense that they are orthogonal to all states vanishing at that stratum. The vanishing condition can be given sense in the framework of holomorphic quantization, where wave functions are true functions and not just classes of functions. In \cite{HRS} we have constructed this costratification for a toy model with gauge group $\SU(2)$ on a single lattice plaquette. As physical effects, we have found a nontrivial overlap between distant strata and, for a certain range of the coupling, a very large transition probability between the ground state of the lattice Hamiltonian and one of the two secondary strata.

In the present paper we deal with the theory with gauge group $G = \SU(2)$ for an arbitrary finite lattice. In that case, there are non-trivial relations characterizing the classical gauge orbit strata which, in a first step, should be implemented on quantum level. This problem has been solved in \cite{FuRS} using the above mentioned holomorphic picture. In this picture, each element $\tau$ of the stratification corresponds to the zero locus of a finite subset $\{p_1, \ldots, p_r\}$ of the algebra $\mc R$ of $G$-invariant representative functions on $G^N_\CC$.\footnote{Here, $G_\CC$ denotes the complexification of $G$.} Viewing the invariants $p_i$ as multiplication operators $\hat p_i$ on the Hilbert space $\mc H$, the union of their images defines a subspace of $\mc H$ whose orthogonal complement $\mc H_\tau$ is, by definition, the element of the costratification corresponding to $\tau$. Thus, to construct $\mc H_\tau$, one has to determine the images of the $\hat p_i$ explicitly. To accomplish that goal, we construct an orthonormal basis in $\mc H$ and determine the structure constants of the algebra $\mc R$ with respect to that basis. This part of our analysis applies to any compact Lie group $G$. So, assuming that we are given the classical stratification for some Lie group $G$ in terms of the classical invariants $p_i$, with the above result at our disposal, 
we can in principle determine the operators  $\hat p_i$  as well as their images in $\mc H$ in terms of linear combinations of the elements of the chosen basis.  

For $G= \SU(2)$, our procedure boils down to a problem in combinatorics of angular momentum theory. For the latter we refer to \cite{BL1,BL2,Louck,Yutsis,Stedman}. Using this theory, 
we obtain the union of the images of the operators  $\hat p_i$ as a subspace of $\mc H$ generated by vectors whose coefficients with respect to our basis are given in terms of Wigner's $3nj$ symbols. The latter are further expressed in terms of $9j$ symbols. For these symbols there exist nowadays efficient calculators, that is, the above coefficients can be calculated explicitly. Using the same techniques, we are also able to reduce the eigenvalue problem for the Hamiltonian of this theory to a problem in linear algebra. 

The paper is organized as follows. In Section \ref{model} we explain the model. To keep the presentation self-contained, in Section \ref{A-SQM} we present 
the basics of stratified quantum gauge theory for arbitrary compact gauge groups as developed in detail in \cite{FuRS}. Moreover, we construct a basis of $\mc H$ consisting of $G$-invariant 
representative functions and derive the multiplication law in $\mc R$. In Section \ref{model-SU2} we turn to the study of the case $G= \SU(2)$. We describe the orbit type strata, analyze the zero loci in terms of the above basis and obtain the images of the multiplication operators $\hat p_i$ in terms of linear combinations of the basis elements. As a consequence, the costrata are given by systems of linear equations with real coefficients built from $3nj$ symbols. We illustrate the result for the special case $N = 2$. Finally, we discuss the eigenvalue problem for the Hamiltonian in terms of the above basis. The text is completed by two appendices containing the proofs of two 
technical results.


\section{The model}
\label{model}


Let $G$ be a compact Lie group and let $\mf g$ be its Lie algebra. Later on, we will specify $G = \SU(2)$, but for the time being, this is not necessary. Let $\Lambda$ be a finite spatial lattice and let $\Lambda^0$, $\Lambda^1$ and $\Lambda^2$ denote, respectively, the sets of lattice sites, lattice links and lattice plaquettes. For the links and plaquettes, let there be chosen an arbitrary orientation. In lattice gauge theory with gauge group $G$ in the Hamiltonian approach, gauge fields (the variables) are approximated by their parallel transporters along links and gauge transformations (the symmetries) are approximated by their values at the lattice sites. Thus, the classical configuration space is the space $G^{\Lambda^1}$ of mappings $\Lambda^1 \to G$, the classical symmetry group is the group $G^{\Lambda^0}$ of mappings $\Lambda^0 \to G$ with pointwise multiplication and the action of $g \in G^{\Lambda^0}$ on $a \in G^{\Lambda^1}$ is given by  
\beq
\label{G-Wir-voll}
(g \cdot a)(\ell) := g(x) a(\ell) g(y)^{-1}\,,
\eeq
where $\ell \in \Lambda^1$ and $x$, $y$ denote the starting point and the endpoint of $\ell$, respectively. The classical phase space is given by the associated Hamiltonian $G$-manifold \cite{AbrahamMarsden,Buch} and the reduced classical phase space is obtained from that by symplectic reduction \cite{OrtegaRatiu,Buch,SjamaarLerman}. We do not need the details here. 
Dynamics is ruled by the classical counterpart of the Kogut-Susskind lattice Hamiltonian, see Subsection \ref{Hamilton}. When identifying $\ctg G$ with $G \times \mf g$, and thus $\ctg G^{\Lambda^1}$ with $G^{\Lambda^1} \times \mf g^{\Lambda^1}$, by means of left-invariant vector fields, the classical Hamiltonian is given by 
\beq
\label{Hamiltonian-C}
H(a,E)
 = 
\frac{g^2}{2 \delta} \sum_{\ell \in \Lambda^1}^N \|E(\ell)\|^2
 -
\frac{1}{g^2 \delta} \sum_{p \in \Lambda^2} \left(\tr a(p) + \ol{\tr a(p)}\right)
 \,,
\eeq
where $a \in G^{\Lambda^1}$, $g$ denotes the coupling constant, $\delta$ denotes the lattice spacing and $a(p)$ denotes the product of $a(\ell)$ along the boundary of the plaquette $p$ in the induced orientation. The trace is taken in some chosen unitary representation. Unitarity ensures that the  Hamiltonian does not depend on the choice of plaquette orientations. Finally, 
$ E \in \mf g^{\Lambda^1}$ is the classical colour electric field (canonically conjugate momentum).

When discussing orbit types in continuum gauge theory, it is convenient to first factorize with respect to the free action of pointed gauge transformations, thus arriving at an action of the compact gauge group $G$ on the quotient manifold. This preliminary reduction can also be carried out in the case of lattice gauge theory under consideration. In fact, given a lattice site $x_0$, it is not hard to see that the normal subgroup 
\beq
\label{G-ptgautrf}
\{g \in G^{\Lambda^0} : g(x_0) = \II\}\,,
\eeq
where $\II$ denotes the unit element of $G$, acts freely on $G^{\Lambda^1}$. Hence, one may pass to the quotient manifold and the residual action by the quotient Lie group of $G^{\Lambda^0}$ with respect to this normal subgroup. Clearly, the quotient Lie group is  naturally isomophic to $G$. The quotient manifold can be identified with a direct product of copies of $G$ and the quotient action can be identified with the action of $G$ by diagonal conjugation as follows. Choose a maximal tree $\mc T$ in the graph $\Lambda^1$ and define the tree gauge of $\mc T$ to be the subset
$$
\{a \in G^{\Lambda^1} : a(\ell) = \II ~ \forall \, \ell \in \mc T\}
$$
of $G^{\Lambda^1}$. One can readily see that every element of $G^{\Lambda^1}$ is conjugate under $G^{\Lambda^0}$ to an element in the tree gauge of $\mc T$ and that two elements in the tree gauge of $\mc T$ are conjugate under $G^{\Lambda^0}$ if they are conjugate under the action of $G$ via constant gauge transformations. This implies that the natural inclusion mapping of the tree gauge into $G^{\Lambda^1}$ descends to a $G$-equivariant diffeomorphism from that tree gauge onto the quotient manifold of $G^{\Lambda^1}$ with respect to the action of the subgroup \eqref{G-ptgautrf}. Finally, by choosing a numbering of the off-tree links in $\Lambda^1$, we can identify the tree gauge of $\mc T$ with the direct product of $N$ copies of $G$, where $N$ denotes the number of off-tree links. This number does not depend on the choice of $\mc T$. Then, the action of $G$ on the tree gauge via constant gauge transformations translates into the action of $G$ on $G^N$ by diagonal conjugation,
\beq\label{G-Wir-Q}
g \cdot (a_1,\dots,a_N) = (g a_1 g^{-1} , \dots , g a_N g^{-1})\,.
\eeq
As a consequence of these considerations, for the discussion of the role of orbit types we may pass from the original large Hamiltonian system with symmetries, given by the configuration space $G^{\Lambda^1}$, the symmetry group $G^{\Lambda^0}$ and the action \eqref{G-Wir-voll}, to the smaller Hamiltonian system with symmetries given by the configuration space 
$$
Q := G^N\,,
$$
the symmetry group $G$ and the action of $G$ on $Q$ given by diagonal conjugation \eqref{G-Wir-Q}. This is the system we will discuss here. As before, the classical phase space is given by the associated Hamiltonian $G$-manifold and the reduced classical phase space is obtained from that by symplectic reduction. One can show that the latter is isomorphic, as a stratified symplectic space, to the reduced classical phase space defined by the original Hamiltonian system with symmetries. 

We will need the following information about the classical phase space. As a space, it is given by the cotangent bundle 
$$
\ctg Q \equiv \ctg G^N\,.
$$
It is a general fact that the action of $G$ on $Q$ naturally lifts to a symplectic action on $\ctg Q$ (consisting of the corresponding 'point transformations' in the language of canonical transformations) and that the lifted action admits a momentum mapping
$$
\mu : \ctg Q \to \mf g^\ast
 \,,~~~~~~
\mu(p)\big(A) := p(A_\ast)\,,
$$
where $p \in \ctg Q$, $A \in \mf g$ and $A_\ast$ denotes the Killing vector field defined by $A$. An easy calculation shows that under the global trivialization
\beq\label{G-trviz-N}
\ctg G^N \cong G^N \times \mf g^N
\eeq
induced by left-invariant vector fields and an invariant scalar product on $\mf g$, the lifted action is given by diagonal conjugation,
\beq\label{G-Wir-P}
g \cdot (a_1,\dots,a_N,A_1,\dots,A_N)
 = 
\big(g a_1 g^{-1} , \dots , g a_N g^{-1} , \Ad(g) A_1 , \dots , \Ad(g) A_N\big)
\eeq
and the associated momentum mapping is given by 
\beq\label{G-ImpAbb}
\mu(a_1,\dots,a_N,A_1,\dots,A_N)
 =
\sum_{i = 1}^N \Ad(a_i) A_i - A_i\,,
\eeq
see e.g.\ \cite[\S 10.7]{Buch}. The reduced phase space $\pha$ is obtained from $\ctg G^N$ by singular symplectic reduction at $\mu = 0$. That is, $\pha$ is the set of orbits of the lifted action of $G$ on the invariant subset $\momap^{-1}(0) \subset \ctg Q$, endowed with the quotient topology induced from the relative topology on this subset. In lattice gauge theory, the condition $\mu=0$ corresponds to the Gau{\ss} law constraint. As a matter of fact, the action of $G$ on $\momap^{-1}(0)$ has the same orbit types as that on $Q$. By definition, the orbit type strata of $\pha$ are the connected components of the subsets of $\pha$ of elements with a fixed orbit type. They are called strata because they provide a stratification of $\pha$ \cite{SjamaarLerman,OrtegaRatiu}. By the procedure of symplectic reduction, the orbit type strata of $\pha$ are endowed with symplectic manifold structures. The bundle projection $\ctg Q \to Q$ induces a mapping $\pha \to Q/G$. This mapping is surjective, because $\momap$ is linear on the fibres of $\ctg Q$ and hence $\momap^{-1}(0)$ contains the zero section of $\ctg Q$. It need not preserve the orbit type though.

\bbm

The tree gauge need not be invariant under time evolution with respect to
a gauge-invariant Hamiltonian (e.g., the Kogut-Susskind lattice Hamiltonian), but
every motion in the full configuration space $G^{\Lambda^1}$ can be transformed
by a time-dependent gauge transformation to the tree gauge. Thus, up to
time-dependent gauge transformations, the tree gauge is invariant under time
evolution. This is reflected in the isomorphism of the reduced phase spaces mentioned above.
\qeb

\ebm


\section{Stratified quantum theory}
\label{A-SQM}


\subsection{Quantization and reduction}
\label{Quantization-Reduction}

To construct the quantum theory of the reduced system, one may either first reduce the classical system and then quantize or first quantize and then reduce the quantum system. Here, we follow the second strategy, that is, we carry out geometric (K\"ahler) quantization on $\ctg G^N$ and subsequent reduction. Let $\mf g_\CC$ denote the complexification of $\mf g$ and let $G_\CC$ denote the complexification of $G$. This is a complex Lie group having $G$ as its maximal compact subgroup. It is unique up to isomorphism. For $G = \SU(n)$, we have $G_\CC = \SL(n,\CC)$. By restriction, the exponential mapping  
$$
\exp : \mf g_\CC \to G_\CC
$$
of $G_\CC$ and multiplication in $G_\CC$ induce a diffeomorphism
\beq\label{G-poldec-1}
G\times\mf g \to G^\CC
 \,,~~~~~~
(a,A) \mapsto a\exp(\mr i A)\,,
\eeq
which is equivariant with respect to the action of $G$ on $G\times \mf g$ by 
$$
g \cdot (a,A) := \big(g a g^{-1} , \Ad(g) A\big)
$$
and the action of $G$ on $G_\CC$ by conjugation. For $G = \SU(n)$, this diffeomorphism amounts to the inverse of the polar decomposition. By applying this diffeomorphism	to each copy, we obtain a diffeomorphism
$$
G^N\times\mf g^N \to G_\CC^N
 \,,~~
(a_1,\dots,a_N , A_1,\dots,A_N)
 \mapsto 
\big(a_1\exp(\mr i A_1),\dots,a_N\exp(\mr i A_N)\big)\,.
$$
By composing the latter with the global trivialization \eqref{G-trviz-N}, we obtain a diffeomorphism 
\beq\label{G-dfm-P}
\ctg G^N \to G_\CC^N
\eeq 
which, due to \eqref{G-Wir-P}, is equivariant with respect to the lifted action of $G$ on $\ctg G^N$ and the action of $G$ on $G_\CC^N$ by diagonal conjugation. Via this diffeomorphism, the complex structure of $G_\CC^N$ and the symplectic structure of $\ctg G^N$ combine to a K\"ahler structure. Half-form K\"ahler quantization on $G_\CC^N$ yields the Hilbert space 
$$
HL^2(G_\CC^N , \mr d \nu_\hbar)
$$
of holomorphic functions on $G_\CC^N$ which are square-integrable with respect to the measure 
\beq
\label{measure-nu}
\mr d \nu_\hbar = \mr e^{- \kappa / \hbar} \, \eta \, \ve\,,
\eeq
where 
$$
\kappa(a_1 \mr e^{\mr i A_1} , \dots , a_N \mr e^{\mr i A_N})
 =
|A_1|^2 + \cdots + |A_N|^2
$$
is the K\"ahler potential on $G_\CC^N$,
$$
\eta(a_1 \mr e^{\mr i A_1} , \dots , a_N \mr e^{\mr i A_N})
 =
\sqrt{\det\frac{\sin\big(\ad(A_1)\big)}{\ad(A_1)}}
 \cdots
\sqrt{\det\frac{\sin\big(\ad(A_N)\big)}{\ad(A_N)}}
$$
is the half-form correction and 
$$
\ve(a_1 \mr e^{\mr i A_1} , \dots , a_N \mr e^{\mr i A_N})
 = 
\mr d a_1 \cdots \mr d a_N \, \mr d A_1 \cdots \mr d A_N
$$
is the Liouville measure on $\ctg G^N$. Reduction then yields the closed subspace 
$$
\mc H = HL^2(G_\CC^N , \mr d \nu_\hbar)^G
$$
of $G$-invariants as the Hilbert space of the reduced system.  

\bbm
\label{Segal-Bargmann}
The above result belongs to Hall \cite{Hall:cptype}. Alternatively, the Hilbert space $HL^2(G_\CC^N , \mr d \nu)$ is obtained via the Segal-Bargmann transformation for compact Lie groups \cite{Hall:SBT}. In more detail, the Segal-Bargmann transformation 
$$
\Phi: L^2 (G^N) \to  HL^2(G_\CC^N , \mr d \nu_\hbar)
$$
is a unitary isomorphism, which restricts to a unitary isomorphism of the subspaces of invariants. 
\qeb

\ebm



\subsection{Orbit type costratification}

Following Huebsch\-mann \cite{Hue:Quantization}, we define the subspaces associated with the orbit type strata of $\pha$ to be the orthogonal complements of the subspaces of functions vanishing on those strata. To accomplish this idea, we first clarify how to interpret elements of $\mc H$ as functions on $\pha$. In the case $N=1$ discussed in \cite{HRS} and \cite{adg}, this is readily done by observing that $\pha \cong T_\CC/W$, where $T$ is a maximal torus in $G$ and $W$ the corresponding Weyl group, and using the isomorphism $HL^2(G_\CC , \mr d \nu)^G \cong HL^2(T_\CC,\mr d \nu_T)^W$, see \S 3.1 of \cite{HRS}. Here, the measure $\mr d \nu_T$ is obtained from $\mr d \nu$ by integration over the conjugation orbits in $G_\CC$, thus yielding an analogue of Weyl's integration formula for $HL^2(G_\CC^N , \mr d \nu)$. In the general case, the argument is as follows. 

First, we construct a quotient of $G_\CC^N$ on which the elements of $\mc H$ define functions. Consider the action of $G_\CC$ on $G_\CC^N$ by diagonal conjugation. For $\ul a \in G_\CC^N$, let $G_\CC \cdot \ul a$ denote the corresponding orbit. Since $G_\CC$ is not compact, $G_\CC \cdot \ul a$ need not be closed. If a holomorphic function on $G_\CC^N$ is invariant under the action of $G$ by diagonal conjugation, then it is also invariant under the action of $G_\CC$ by diagonal conjugation, i.e., it is constant on the orbit $G_\CC \cdot \ul a$ for every $\ul a \in G_\CC^N$. Being continuous, it is then constant on the closure $\ol{G_\CC \cdot \ul a}$. As a consequence, it takes the same value on two orbits whenever their closures intersect. This motivates the following definition. Two elements $\ul a , \ul b \in G_\CC^N$ are said to be orbit closure equivalent if there exist $\ul c_1 , \dots , \ul c_r \in G_\CC^N$ such that
$$
\ol{G_\CC \cdot \ul a} \cap \ol{G_\CC \cdot \ul c_1} \neq \varnothing
 \,,~~~
\ol{G_\CC \cdot \ul c_1} \cap \ol{G_\CC \cdot \ul c_2} \neq \varnothing
 \,,~~~ \dots \,,~~~
\ol{G_\CC \cdot \ul c}_r \cap \ol{G_\CC \cdot \ul b} \neq \varnothing\,.
$$
Clearly, orbit closure equivalence defines an equivalence relation on $G_\CC^N$, indeed. Let $G_\CC^N /\!/ G_\CC$ denote the topological quotient\footnote{This notation is motivated by the fact that the quotient provides a categorical quotient of $G_\CC^N$ by $G_\CC$ in the sense of geometric invariant theory \cite{Mumford}.}. By construction, the elements of $\mc H$ descend to continuous functions on $G_\CC^N /\!/ G_\CC$. 

In \cite{FuRS} we have explained in some detail how the orbit closure quotient $G_\CC^N /\!/ G_\CC$ is related to the reduced phase space $\pha$. This discussion is based on the observation that, 
via the equivariant diffeomorphism \eqref{G-dfm-P}, the momentum mapping may be viewed as a mapping
$$
\mu : G_\CC^N \to \mf g^\ast
$$
and, thus, $\pha$ may be viewed as the quotient of $\mu^{-1}(0) \subset G_\CC^N$ by the action of $G$. In this language,  $\mu^{-1}(0)$ turns out to be a Kempf-Ness set \cite{KempfNess}. 
Using this fact, one can prove the following. 

\btm\label{T-KN}

The natural inclusion mapping $\mu^{-1}(0) \to G_\CC^N$ induces a homeomorphism 
\beq\label{G-KN-Hoem}
\pha \to G_\CC^N /\!/ G_\CC\,.
\eeq

\etm

For the proof, see \cite{HeinznerLoose}.

As a by-product of the proof, one finds that two points $\ul a , \ul b \in G_\CC^N$ are orbit closure equivalent iff 
\beq\label{G-KN-rel}
\ol{G_\CC \cdot \ul a} \cap \ol{G_\CC \cdot \ul b} \cap \mu^{-1}(0) \neq \varnothing\,.
\eeq
As a result, via the homeomorphism \eqref{G-KN-Hoem}, the elements of $\mc H$ can be interpreted as functions on $\pha$. By virtue of this interpretation, to a given orbit type stratum $\pha_\tau \subset \pha$, there corresponds the closed subspace 
$$
\mc V_{\tau} := \{\psi \in \mc H : \psi_{\res \pha_\tau} = 0\}\,.
$$
We define the subspace $\mc H_\tau$ associated with $\pha_\tau$ to be the orthogonal complement of $\mc V_\tau$ in $\mc H$. Then, we have the orthogonal decomposition
$$
\mc H_\tau \oplus \mc V_\tau = \mc H\,.
$$

\bbm\label{Bem-HS}

Since holomorphic functions are continuous, one has 
\beq\label{G-D-V}
\mc V_{\tau} = \{\psi \in \mc H : \psi_{\res \ol{\pha_\tau}} = 0\}\,.
\eeq
First, since the principal stratum is dense in $\pha$, this implies that the
subspace associated with that stratum coincides with $\mc H$. Thus, in the
discussion of the orbit type subspaces below, the principal stratum may be
ignored. Second, recall that in a stratification, the strata satisfy the
condition of the frontier, which means that if $\pha_{\sigma} \cap
\ol{\pha_{\tau}} \neq \varnothing$, then $\pha_{\sigma} \subset \ol{\pha_{\tau}}$.
In view of this, \eqref{G-D-V} implies that if $\pha_{\sigma} \cap
\ol{\pha_{\tau}} \neq \varnothing$, then $\mc V_\tau \subset \mc V_\sigma$ and
hence $\mc H_\sigma \subset \mc H_\tau$. The family of orthogonal projections  
$$
\mc H_\tau \to \mc H_\sigma
 \text{ whenever } 
\pha_{\sigma} \cap \ol{\pha_{\tau}} \neq \varnothing
$$
makes the family of closed subspaces $\mc H_\tau$ into a costratification in the
sense of Huebsch\-mann \cite{Hue:Quantization}. 
\qeb

\ebm

In order to analyse the condition $\psi_{\res \pha_\tau} = 0$, it is convenient to work with those subsets of $G_\CC^N$ which under the natural projection $G_\CC^N \to G_\CC^N /\!/ G_\CC$ and the homeomorphism \eqref{G-KN-Hoem} correspond to the orbit type strata of $\pha$. For a given orbit type stratum $\pha_\tau$, denote this subset by $(G_\CC^N)_\tau$. That is, $(G_\CC^N)_\tau$ consists of the elements $\ul a$ of $G_\CC^N$ whose orbit closure equivalence class belongs to the image of $\pha_\tau$ under the homeomorphism \eqref{G-KN-Hoem}. In other words, $\ul a \in (G_\CC^N)_\tau$ iff it is orbit closure equivalent to some element of $\mu^{-1}(0)$ whose $G$-orbit belongs to $\pha_\tau$. Clearly,
\beq\label{G-V-GC}
\mc V_\tau = \{\psi \in \mc H : \psi_{\res (G_\CC^N)_\tau} = 0\}\,.
\eeq


\subsection{Characterization of costrata in terms of relations}
\label{A-SQT-AA-defRel}


To conclude the general discussion, we describe how to construct $\mc V_\tau$ and $\mc H_\tau$ using defining relations for the orbit type strata $\pha_\tau$.

Let $\mf R(G^N)$ denote the commutative algebra of representative functions on $G^N$ and let $\mc R := \mf R(G^N)^G$ be the subalgebra of $G$-invariant elements.  
Since $G_\CC^N$ is the complexification of the compact Lie group $G^N$, the proposition and Theorem 3  in Section 8.7.2 of \cite{Procesi:LG} imply that $\mf R(G^N)$ coincides with the coordinate ring of $G_\CC^N$, viewed as a complex affine variety, and that $\mf R(G^N)$ coincides with the algebra of representative functions on $G^N_\CC$. 
As a consequence, $\mc R$ coincides with the algebra of $G$-invariant representative functions on $G^N_\CC$, where the relation is given by restriction and analytic continuation, respectively.  

Recall that an ideal $\mc I \subset \mc R$ is called a radical ideal if for all $f \in \mc R$ satisfying $f^n \in \mc I$ for some $n$ one has $f \in \mc I$. Moreover, given a subset 
$R \subset \mc R$, one defines the zero locus of $R$ by 
$$
\{\ul a \in G^N_\CC : f(\ul a) = 0 \text{ for all } f \in R\} \subset G^N_\CC\,.
$$
It coincides with the zero locus of the ideal in $\mc R$ generated by $R$.

\bsz
\label{S-V}

Let $\pha_\tau$ be an orbit type stratum and let $R_\tau$ be a subset of $\mc R$ satisfying

\ben

\sitem\label{i-S-V-Loc}
The zero locus of $R_\tau$ coincides with the topological closure of $(G_\CC^N)_\tau$,

\sitem\label{i-S-V-Rad}
The ideal generated by $R_\tau$ in $\mc R$ is a radical ideal.

\een

Then, $\mc V_\tau$ is obtained by intersecting $\mc H$ with the ideal generated algebraically by $R_\tau$ in the algebra $\text{\rm Hol}(G^N_\CC)^G$ of $G$-invariant 
holomorphic functions on $G_\CC^N$.

\esz

For the proof, see \cite{FuRS}.

By Hilbert's Basissatz, finite subsets $R_\tau \subset \mc R$ satisfying conditions \rref{i-S-V-Loc} and \rref{i-S-V-Rad} of Proposition \rref{S-V} exist. Given $R_\tau$, Proposition \rref{S-V} implies the following explicit characterization of the subspaces $\mc V_\tau$ and $\mc H_\tau$ in terms of multiplication operators. For $f \in \mc R$, let $\hat f : \mc H \to \mc H$ denote the operator of multiplication by $f$.

\bfg\label{F-V}

Let $\pha_\tau$ be an orbit type stratum and let $R_\tau = \{p_1 , \dots , p_r\}$ be a finite subset of $\mc R$ satisfying conditions \rref{i-S-V-Loc} and \rref{i-S-V-Rad} 
of Proposition \rref{S-V}. Then,
\eqqedan
\beq
\mc V_\tau = \im(\hat p_1) + \cdots + \im(\hat p_r)
 \,,\qquad
\mc H_\tau = \ker\big(\hat p_1^\dagger\big) \cap \cdots \cap \ker\big(\hat p_r^\dagger\big)\,.
\eqqed
\eeq
\eqqedaus

\efg

In what follows, we will refer to conditions \rref{i-S-V-Loc} and \rref{i-S-V-Rad} of Proposition \rref{S-V} as the zero locus condition and the radical ideal condition, respectively.



\subsection{The commutative algebra $\mc R$}
\label{A-reprF}


By Proposition \ref{S-V} and Corollary \ref{F-V}, the costratification of the quantum Hilbert space $\mc H$ is given by a family of finite subsets $R_\tau \subset \mc R$ satisfying 
conditions \rref{i-S-V-Loc} and \rref{i-S-V-Rad}. Each of these subsets consists of a finite set of $G$-invariant polynomials on $G^N_\CC$. To construct the costratification explicitly, one has to find the images of the multiplication operators defined by these invariant polynomials. This can be achieved by choosing an orthonormal basis in $\mc H$ and by finding the structure constants of the multiplication law in $\mc R$ in that basis.

By Remark \ref{Segal-Bargmann}, we can first consider the Hilbert space $ L^2 (G^N)^G$ and use the theory of compact Lie groups. For the convenience of the reader, and to fix the notation, we  recall some basics, see e.g.\ \cite{Naimark} or \cite{Goodman} for details. Below, all representations are assumed to be continuous and unitary without further notice. Let $\widehat G$ denote the set of isomorphism classes of finite-dimensional irreps of $G$. Given a finite-dimensional unitary representation $(H, \pi)$ of $G$, let $C(G)_\pi \subset \mf R(G)$ denote the subspace of representative functions\footnote{The subspace spanned by all matrix coefficients $\langle \zeta , \pi(\cdot) v \rangle $ with $v \in H$ and $\zeta \in H^\ast$.} of $\pi$ and 
let $\chi_\pi \in C(G)_\pi$ be the character of $\pi$, defined by $\chi_\pi(a) := \tr\big(\pi(a)\big)$. The same notation will be used for the Lie group $G^N$.

The elements of $\widehat G$ will be labeled by the corresponding highest weight $\lambda$ relative to some chosen Cartan subalgebra and some chosen dominant Weyl chamber. Assume that for every $\lambda \in \widehat G$ a concrete unitary irrep $(H_\lambda,\pi_\lambda)$ of highest weight $\lambda$ in the Hilbert space $H_\lambda$ has been chosen. Given $\ul\lambda = (\lambda^1,\dots,\lambda^N) \in \widehat G^N$, we define a representation $(H_{\ul\lambda},\pi_{\ul\lambda})$ of $G^N$ by 
\beq
\label{G-irrepsGN}
H_{\ul\lambda} = \bigotimes_{i = 1}^N H_{\lambda^i}
\,,\quad 
\pi_{\ul\lambda}(\ul a) = \bigotimes_{i=1}^N \pi_{\lambda^i}(a_i)\,,
\eeq
where $\ul a = (a_1 , \dots , a_N)$. This representation is irreducible and we have 
$$
C(G^N)_{\pi_{\ul\lambda}} \cong \bigotimes_{i = 1}^N C(G)_{\pi_{\lambda^i}}
\,,
$$ 
isometrically with respect to the $L^2$-norms. Using this, together with the Peter-Weyl theorem for $G$, we obtain that $\bigoplus_{\ul\lambda \in \widehat G^N} C(G^N)_{\pi_{\ul\lambda}}$ is dense in $L^2 (G^N, {\rm d}^N a)$. Since $\bigoplus_{\ul\lambda \in \widehat G^N} C(G^N)_{\pi_{\ul\lambda}} \subset \bigoplus_{\pi \in \widehat{G^N}} C(G^N)_\pi$, this implies

\ble\label{L-ProdReps}

Every irreducible representation of $G^N$ is equivalent to a product representation $(H_{\ul\lambda},\pi_{\ul\lambda})$ with $\ul\lambda \in \widehat G^N$. If $(H_{\ul\lambda},\pi_{\ul\lambda})$ and $(H_{\ul\lambda'},\pi_{\ul\lambda'})$ are isomorphic, then $\ul\lambda = \ul\lambda'$.
\qed

\ele

Given $\ul\lambda \in \widehat G^N$, let $\pi^d_{\ul\lambda}$ denote the representation of $G$ on $H_{\ul\lambda}$ defined by  
\beq\label{pi-d}
\pi^d_{\ul\lambda}(a) := \pi_{\ul\lambda} (a, \ldots, a)
\,.
\eeq
This representation will be referred to as the diagonal representation induced by $\pi_{\ul\lambda}$. It is reducible and has the isotypical decomposition
$$
H_{\ul\lambda}
 =
\bigoplus_{\lambda \in \widehat G} H_{\ul\lambda,\lambda}
$$
into uniquely determined subspaces $H_{\ul\lambda,\lambda}$. Recall that these subspaces may be obtained as the images of the orthogonal projectors 
\beq
\label{Proj-irrep}
{\mathbb P}_\lambda
 := 
\dim (H_\lambda) \int_G \ol{\chi_{\pi_\lambda}(a)} \, \pi_{\ul\lambda}(a) \, \mr d a 
\eeq
on $H_{\ul\lambda}$. These projectors commute with one another and with $\pi^d_{\ul\lambda}$. If an isotypical subspace $H_{\ul\lambda,\lambda}$ is reducible, we can further decompose it in a non-unique way into irreducible subspaces of isomorphism type $\lambda$. Let $m_{\ul\lambda}(\lambda)$ denote the number of these irreducible subspaces (the multiplicity of $\pi_\lambda$ in $\pi^d_{\ul\lambda}$) and let $\widehat G_{\ul\lambda}$ denote the subset of $\widehat G$ consisting of the highest weights $\lambda$ such that $m_{\ul\lambda}(\lambda) > 0$. This way, we obtain a unitary $G$-representation isomorphism 
\beq\label{G-D-vp}
\vp_{\ul\lambda}
 : 
H_{\ul\lambda}
 ~\to~
\bigoplus_{\lambda \in \widehat G_{\ul\lambda}} 
 \,
\bigoplus_{k=1}^{m_{\ul\lambda}(\lambda)} 
H_\lambda
\,.
\eeq
Let 
$$
\pr^{\ul\lambda}_{\lambda,k}
 : 
\bigoplus_{\lambda \in \widehat G_{\ul\lambda}} 
 \,
\bigoplus_{k=1}^{m_{\ul\lambda}(\lambda)} H_\lambda
 \to 
H_\lambda
 \,,\qquad
\mr i^{\ul\lambda}_{\lambda,k}
 : 
H_\lambda
 \to 
\bigoplus_{\lambda \in \widehat G_{\ul\lambda}} 
 \,
\bigoplus_{k=1}^{m_{\ul\lambda}(\lambda)} H_\lambda \, ,
$$
denote the natural projections and injections of the direct sum. For every $\lambda \in \widehat G_{\ul\lambda}$ and every $k,l = 1, \dots , m_{\ul\lambda}(\lambda)$, define a $G$-representation endomorphism $A^{\ul\lambda,\lambda}_{k,l}$ of $\pi^d_{\ul\lambda}$ by
\beq
\label{A-T}
A^{\ul\lambda,\lambda}_{k,l}
 := 
\frac{1}{\sqrt{\dim (H_\lambda)}}
 ~
\vp^{-1}_{\ul\lambda}
 \circ 
\mr i^{\ul\lambda}_{\lambda,k} \circ \pr^{\ul\lambda}_{\lambda,l}
 \circ 
\vp_{\ul\lambda}
\eeq
and a $G$-invariant function $\BF{\ul\lambda}{\lambda}{k,l}$ on $G^N$ by 
\beq
\label{ReprF-Hom}
\BF{\ul\lambda}{\lambda}{k,l}(\ul a)
 := 
\sqrt{\dim(H_{\ul\lambda})}
\,
\tr\left(\pi_{\ul\lambda}(\ul a) A^{\ul\lambda,\lambda}_{k,l}\right)
\,.
\eeq

\bsz\label{S-ReprF}

The family of functions 
$$
\left\{
\BF{\ul\lambda}{\lambda}{k,l}
~:~
\ul\lambda \in \widehat G^N
,~
\lambda \in \widehat G_{\ul\lambda}
\,,~
k,l = 1 , \dots , m_{\ul\lambda}(\lambda)
\right\}
$$
constitutes an orthonormal basis in  $L^2(G^N)^G$.

\esz

\bbw

Note that for every $\ul\lambda \in \widehat G^N$, the mapping 
\beq
\label{EquIso}
T_{\ul\lambda} : \End(H_{\ul\lambda}) \to C(G^N)_{\pi_{\ul\lambda}}
 \,,\qquad 
T_{\ul\lambda}(A)(a)
 := 
\sqrt{\dim H_{\ul\lambda}} \, \tr\big(\pi_{\ul\lambda}(a) A\big)
\,,
\eeq
is a unitary $G$-representation isomorphism with respect to the scalar product on $\End(H_{\ul\lambda})$ defined by $\braket A B = \tr(A^\ast B)$ and the induced endomorphism representation on $\End(H_{\ul\lambda})$, given by assigning to $g \in G$ the automorphism 
$$
A \mapsto \pi^d_{\ul\lambda}(g) \, A \, \pi^d_{\ul\lambda}(g)^{-1}
$$
of $\End(H)$. Being a representation isomorphism, $T_{\ul\lambda}$ restricts to a unitary Hilbert space isomorphism of the subspaces of $G$-invariant elements, $\End(H_{\ul\lambda})^G \to C(G^N)_{\pi_{\ul\lambda}}^G$. Now, $\End(H_{\ul\lambda})^G$ consists precisely of the representation endomorphisms of $\pi^d_{\ul\lambda}$. Hence, Schur's lemma implies that it is spanned by the endomorphisms $A^{\ul\lambda,\lambda}_{k,l}$ with $\lambda \in \widehat G_{\ul\lambda}$ and $k,l = 1 , \dots , m_{\ul\lambda}(\lambda)$. Using $\pr^{\ul\lambda}_{\lambda,k} \circ \mr i^{\ul\lambda}_{\lambda',k'} = \delta_{\lambda\lambda'} \delta_{kk'} \id_{H_\lambda}$, we compute
$$
\big\langle 
A^{\ul\lambda,\lambda}_{k,l} \big| A^{\ul\lambda,\lambda'}_{k',l'} 
\big\rangle
 = 
\delta_{\lambda\lambda'} \, \delta_{kk'} \, \delta_{ll'}
\,.
$$
It follows that the endomorphisms $A^{\ul\lambda,\lambda}_{k,l}$ with $\lambda \in \widehat G_{\ul\lambda}$ and $k,l = 1 , \dots , m_{\ul\lambda}(\lambda)$ form an orthonormal basis in $\End(H_{\ul\lambda})^G$, and hence that their images under $T_{\ul\lambda}$, i.e., the functions $\BF{\ul\lambda}{\lambda}{k,l}$, form an orthonormal basis in $C(G^N)_{\pi_{\ul\lambda}}^G$. Thus, the family given in the proposition yields an orthonormal basis in $\mc R = \mf R(G^N)^G$. 

It remains to show that $\mc R$ is dense in $L^2(G^N,\mr d^N a)^G$. This follows from the Peter-Weyl theorem for $G^N$ by applying the averaging operator
$$
\PP_G : L^2(G^N,\mr d^N a) \to L^2(G^N,\mr d^N a)^G
 \,,~
\PP_G(f)(\ul a)
 =
\int_G f(g a_1 g^{-1} , \dots , g a_N g^{-1}) \, \mr d g
\,,
$$
and observing that the image of a dense subset under a surjective continuous mapping is dense.
\ebw

By analytic continuation, the irreps $\pi_\lambda$ of $G$ induce irreps $\pi^\CC_\lambda$ of $G_\CC$, the irreps $\pi_{\ul\lambda}$ of $G^N$ induce irreps $\pi^\CC_{\ul\lambda}$ of $G^N_\CC$, and the functions $\BF{\ul\lambda}{\lambda}{k,l}$ on $G^N$ induce holomorphic functions $\BFC{\ul\lambda}{\lambda}{k,l}$ on $G^N_\CC$. Then, \eqref{G-irrepsGN}, \eqref{pi-d} and \eqref{ReprF-Hom} hold with $\pi_{\ul\lambda}$, $\pi_\lambda$ and $\BF{\ul\lambda}{\lambda}{k,l}$ replaced by, respectively, $\pi_{\ul\lambda}^\CC$, $\pi_\lambda^\CC$ and $\BFC{\ul\lambda}{\lambda}{k,l}$.

\bfg\label{F-ReprF}

The family of functions
$$
\left\{
\BFC{\ul\lambda}{\lambda}{k,l}
~:~
\ul\lambda \in \widehat G^N
,~
\lambda \in \widehat G_{\ul\lambda}
\,,~
k,l = 1 , \dots , m_{\ul\lambda}(\lambda)
\right\}
$$
constitutes an orthogonal basis in $\mc H$. The norms are
\beq
\label{norm}
\|\BFC{\ul\lambda}{\lambda}{k,l}\|^2
 =
\prod_{r=1}^N C_{\lambda^r}
 \,,\qquad
C_{\lambda^r} = (\hbar\pi)^{\dim(\group)/2}\mr e^{\hbar|\lambda^r+\rho|^2},
\eeq
where $\rho$ denotes half the sum of the positive roots. The expansion coefficients of $f \in \mc H$ wrt.\ this basis are given by the scalar products $\braket{\BF{\ul\lambda}{\lambda}{k,l}}{f_{\res G^N}}$ in $L^2(G^N)^G$.

\efg

\bbw

See Appendix \ref{Cor}. The last statement follows from the fact that two elements of $\mc H$ coincide iff their restrictions to $G^N$ coincide. Since the functions $\BF{\ul\lambda}{\lambda}{k,l}$ form an orthonormal basis in $L^2(G^N)^G$, we have $f_{\res G^N} = \sum_{\ul \lambda,\lambda,k,l} \braket{\BF{\ul\lambda}{\lambda}{k,l}}{f_{\res G^N}} \,\, \BF{\ul\lambda}{\lambda}{k,l}$. 
Since $\BF{\ul\lambda}{\lambda}{k,l} = \BFC{\ul\lambda}{\lambda}{k,l}{}_{\res G^N}$, this yields the assertion.
\ebw

\bbm\label{Spin-network}

The orthonormal basis of invariant representative functions provided by Proposition \ref{S-ReprF} is a special case of a spin network basis in the 
sense of Baez \cite{Baez}. It is special in so far as from the very beginning we have fixed a tree gauge, which reduces the group of local gauge transformations 
to the action of $G$.  Moroever, our basis above clearly corresponds to a fixed graph (a finite regular cubic lattice).  In this situation, we are able to provide 
a more explicit presentation of the basis elements in terms of appropriate functions. We refer to \cite{Baez} for comments on various applications 
of spin networks in Mathematical Physics. In particular, over the years spin network states have become an important tool in Loop Quantum Gravity, see \cite{Thiemann} and further references therein. 
\qeb

\ebm
\bigskip

Now, let us turn to the discussion of the multiplication structure of the $G$-invariant representative functions $\BFC{\ul\lambda}{\lambda}{k,l}$. We assume that a unitary $G$-representation isomorphism \eqref{G-D-vp} has been chosen for every $\ul\lambda \in \widehat G^N$ and every $N$. Denote
$$
d_{\lambda} := \dim H_\lambda
 \,,\qquad
d_{\ul \lambda} := \dim H_{\ul \lambda}
\,.
$$
Writing 
 \al{\nonumber
\BFC{\ul\lambda_1}{\lambda_1}{k_1,l_1}(\ul a)&
 \,
\BFC{\ul\lambda_2}{\lambda_2}{k_2,l_2}(\ul a)
\\ \label{G-MF-1}
 & =
\sqrt{d_{\ul\lambda_1} d_{\ul\lambda_2}}
 \,
\tr
 \left(
 \left(
A^{\ul\lambda_1,\lambda_1}_{k_1,l_1} \otimes A^{\ul\lambda_2,\lambda_2}_{k_2,l_2}
 \right)
 \circ
\Big(\pi_{\ul\lambda_1}(\ul a) \otimes \pi_{\ul\lambda_2}(\ul a)\Big)
 \right)
\,,
 }
we see that in order to expand the product $\BFC{\ul\lambda_1}{\lambda_1}{k_1,l_1} \cdot \BFC{\ul\lambda_2}{\lambda_2}{k_2,l_2}$ in terms of the basis functions $\BFC{\ul\lambda}{\lambda}{k,l}$, a reasonable strategy is to decompose the $G^N$-representation $\pi_{\ul\lambda_1} \otimes \pi_{\ul\lambda_2}$ into $G^N$-irreps $\ul\lambda$ and then relate these $G^N$-irreps to the basis functions using the chosen $G$-representation isomorphisms $\vp_{\ul\lambda}$. To implement this, we define two different unitary $G$-representation isomorphisms of the diagonal representation $\pi_{\ul\lambda_1}^d \otimes\pi_{\ul\lambda_2}^d$ with an orthogonal direct sum of $G$-irreps. The first one, $\Phi_{\ul\lambda_1\ul\lambda_2}$, is adapted to the tensor product on the right hand side of \eqref{G-MF-1}. It is defined by 
 \al{\label{eq: decomp11}
\Phi_{\ul\lambda_1\ul\lambda_2}
 : 
H_{\ul\lambda_1}\otimes H_{\ul\lambda_2}
 & \overset{\vp_{\ul\lambda_1}\otimes\vp_{\ul\lambda_2}}{\longrightarrow}
\bigoplus_{\lambda_1,\lambda_2}
 \left(
\bigoplus_{i_1=1}^{m_{\ul \lambda_1}(\lambda_1)}\bigoplus_{i_2=1}^{m_{\ul\lambda_2}(\lambda_2)}\left(H_{\lambda_1}\otimes H_{\lambda_2}\right)
 \right)
\\ \label{eq: decomp12}
  & \hspace{1cm}
\overset{\phi_{\ul\lambda_1\ul\lambda_2}}{\longrightarrow}\bigoplus_{\lambda_1,\lambda_2}
 \left(
\bigoplus_{i_1=1}^{m_{\ul \lambda_1}(\lambda_1)}\bigoplus_{i_2=1}^{m_{\ul\lambda_2}(\lambda_2)}\left(\bigoplus_\lambda\bigoplus_{i=1}^{m_{\lambda_1,\lambda_2}(\lambda)}H_\lambda\right)
 \right)
\,,
 }
where $\phi_{\ul\lambda_1\ul\lambda_2}$ acts on each summand $H_{\lambda_1}\otimes H_{\lambda_2}$ as $\vp_{(\lambda_1,\lambda_2)}$. Let $\pr^{\ul\lambda_1,\ul\lambda_2}_{\lambda_1,\lambda_2,i_1,i_2,\lambda,i}$ and $\mr{i}^{\ul\lambda_1,\ul\lambda_2}_{\lambda_1,\lambda_2,i_1,i_2,\lambda,i}$ denote the natural projection and injection operators of the direct sum \eqref{eq: decomp12}. We have
 \al{\label{eq: prof-inj-z1-1}
\vp_{(\lambda_1,\lambda_2)}
 \circ 
 \left(
\pr_{\lambda_1,i_1}^{\ul\lambda_1}\otimes\pr_{\lambda_2,i_2}^{\ul\lambda_2}
 \right)
 & =
 \left(
\sum_\lambda \sum_{i=1}^{m_{(\lambda_1,\lambda_2)}(\lambda)} 
\mr{i}_{\lambda,i}^{(\lambda_1,\lambda_2)}
 \circ 
\pr^{\ul\lambda_1,\ul\lambda_2}_{\lambda_1,\lambda_2,i_1,i_2,\lambda,i}
 \right)
 \circ 
\phi_{\ul\lambda_1\ul\lambda_2}
\\ \label{eq: prof-inj-z1-2}
\phi_{\ul\lambda_1\ul\lambda_2}
 \circ 
 \left(
\mr{i}^{\ul\lambda_1}_{\lambda_1,i_1}\otimes\mr{i}^{\ul\lambda_2}_{\lambda_2,i_2}
 \right)
 & =
 \left(
\sum_\lambda \sum_{i=1}^{m_{(\lambda_1,\lambda_2)}(\lambda)}
\mr{i}^{\ul\lambda_1,\ul\lambda_2}_{\lambda_1,\lambda_2,i_1,i_2,\lambda,i}
 \circ
\pr^{(\lambda_1,\lambda_2)}_{\lambda,i}
 \right)
 \circ
\vp_{(\lambda_1\lambda_2)}
\,.
 }
The second unitary $G$-representation isomorphism, $\Psi_{\ul\lambda_1\ul\lambda_2}$, is adapted to the definition of the basis functions $\BF{\ul\lambda}{\lambda}{k,l}$. It is defined by
 \al{\label{eq: decomp21}
\Psi_{\ul\lambda_1\ul\lambda_2}
 : 
H_{\ul\lambda_1} \otimes H_{\ul\lambda_2}
 & \overset{\psi_{\ul\lambda_1\ul\lambda_2}^1}{\longrightarrow}
\bigoplus_{\ul \lambda}
 \left(
\bigoplus_{i = 1}^{m_{\ul\lambda_1,\ul\lambda_2}(\ul\lambda)} H_{\ul\lambda}
 \right)
\\ \label{eq: decomp22}
  & \overset{\psi_{\ul\lambda_1\ul\lambda_2}^2}{\longrightarrow}
\bigoplus_{\ul\lambda}
 \left(
\bigoplus_{i=1}^{m_{\ul\lambda_1,\ul\lambda_2}(\ul\lambda)}
 \left(
\bigoplus_\lambda \bigoplus_{k=1}^{m_{\ul\lambda}(\lambda)} H_\lambda
 \right)
 \right)
\,,
 }
where $\psi^1_{\ul\lambda_1\ul\lambda_2}$ is some unitary $G^N$-representation isomorphism, provided by Lemma \ref{L-ProdReps}, and $\psi^2_{\ul\lambda_1\ul\lambda_2}$ is the $G$-representation isomorphism acting on each summand $H_{\ul\lambda}$ as $\vp_{\ul\lambda}$. Moreover, $m_{\ul\lambda_1,\ul\lambda_2}(\ul\lambda)$ is the multiplicity of the $G^N$-irrep $H_{\ul\lambda}$ in 
$H_{\ul\lambda_1} \otimes H_{\ul\lambda_2}$.
Let $\pr^{\ul\lambda_1,\ul\lambda_2}_{\ul\lambda,i}$ and $\mr{i}^{\ul\lambda_1,\ul\lambda_2}_{\ul\lambda,i}$ be the natural projections and injections, respectively, of the direct sum \eqref{eq: decomp21} and let $\pr^{\ul\lambda_1,\ul\lambda_2}_{\ul\lambda,i,\lambda,k}$ and $\mr{i}^{\ul\lambda_1,\ul\lambda_2}_{\ul\lambda,i,\lambda,k}$ be the natural projections and injections, respectively, of the direct sum \eqref{eq: decomp22}. We have
 \al{\label{eq: rel21}
\psi_{\ul\lambda_1\ul\lambda_2}^2
 \circ 
\mr{i}^{\ul\lambda_1,\ul\lambda_2}_{\ul\lambda,i}
 & =
 \left(
\sum_\lambda \sum_{k=1}^{m_{\ul\lambda}(\lambda)}
\mr{i}^{\ul\lambda_1,\ul\lambda_2}_{\ul\lambda,i,\lambda,k} 
 \circ 
\pr^{\ul\lambda}_{\lambda,k}
 \right)
 \circ
\vp_{\ul\lambda}
\,,
\\ \label{eq: rel22}
\vp_{\ul\lambda} \circ \pr^{\ul\lambda_1,\ul\lambda_2}_{\ul\lambda,i}
 & =
 \left(
\sum_{\lambda} \sum_{k=1}^{m_{\ul\lambda}(\lambda)}
\mr{i}^{\ul\lambda}_{\lambda,k}
 \circ 
\pr^{\ul\lambda_1,\ul\lambda_2}_{\ul\lambda,i,\lambda,k}
 \right)
 \circ
\psi^2_{\ul\lambda_1\ul\lambda_2}
\,.
 }
By construction, $\Psi_{\ul\lambda_1\ul\lambda_2} \circ \Phi_{\ul\lambda_1\ul\lambda_2}^{-1}$ is a unitary automorphism of a direct sum of $G$-irreps $H_\lambda$. Hence, Schur's lemma implies that 
 \al{\label{eq: UnitaryCoeff-1}
\pr^{\ul\lambda_1,\ul\lambda_2}_{\ul\lambda,i,\lambda,k}
 \circ
\left(\Psi_{\ul\lambda_1,\ul\lambda_2} \circ \Phi_{\ul\lambda_1,\ul\lambda_2}^{-1}\right)
 \circ
\mr{i}^{\ul\lambda_1,\ul\lambda_2}_{\lambda_1,\lambda_2,i_1,i_2,\lambda',i'}
 =
\delta_{\lambda\lambda'}
 \,
U^{\ul\lambda_1,\ul\lambda_2;\ul\lambda,i,\lambda,k}_{\lambda_1,\lambda_2,i_1,i_2,i'}
 \,
\id_{H_\lambda}
\,,
\\ \label{eq: UnitaryCoeff-2}
\pr^{\ul\lambda_1,\ul\lambda_2}_{\lambda_1,\lambda_2,i_1,i_2,\lambda',i'}
 \circ
\left(\Phi_{\ul\lambda_1\ul\lambda_2} \circ \Psi_{\ul\lambda_1\ul\lambda_2}^{-1}\right)
 \circ
\mr{i}^{\ul\lambda_1,\ul\lambda_2}_{\ul\lambda,i,\lambda,k}
 =
\delta_{\lambda\lambda'}
 \, 
\ol{U^{\ul\lambda_1,\ul\lambda_2;\ul\lambda,i,\lambda,k}_{\lambda_1,\lambda_2,i_1,i_2,i'}}
 \,
\id_{H_\lambda} \, ,
 } 
with certain coefficients $U^{\ul\lambda_1,\ul\lambda_2;\ul\lambda,i,\lambda,k}_{\lambda_1,\lambda_2,i_1,i_2,i'}$.

\bsz\label{MultLaw-Gen}

In terms of the basis functions, the multiplication in $\mc R$ is given by 
 \ala{
&  \BFC{\ul\lambda_1}{\lambda_1}{k_1,l_1} \cdot \BFC{\ul\lambda_2}{\lambda_2}{k_2,l_2}
\\
  = &{\textstyle
\sqrt{\frac{d_{\ul\lambda_1} d_{\ul\lambda_2}}{d_{\lambda_1} d_{\lambda_2}}}}
 \,
\sum_{\ul\lambda} 
 \,
\sum_{n=1}^{m_{\ul\lambda_1,\ul\lambda_2}(\ul\lambda)}
 \,
\sum_\lambda
 \,
\sum_{k,l=1}^{m_{\ul\lambda}(\lambda)}
 \,
\sum_{j=1}^{m_{(\lambda_1,\lambda_2)}(\lambda)}
 \!
\textstyle{\sqrt{\frac{d_\lambda}{d_{\ul\lambda}}}}
 \,
U^{\ul\lambda_1,\ul\lambda_2;\ul\lambda,n,\lambda,k}_{\lambda_1,\lambda_2,k_1,k_2,j}
 ~
\ol{U^{\ul\lambda_1,\ul\lambda_2;\ul\lambda,n,\lambda,l}_{\lambda_1,\lambda_2,l_1,l_2,j}}
 ~
\BFC{\ul\lambda}{\lambda}{k,l}
\,.
 }

\esz

The same formula holds true for the basis functions $\BF{\ul\lambda}{\lambda}{k,l}$ on $G^N$.

\bbw

It suffices to prove the assertion for the basis functions $\BF{\ul\lambda}{\lambda}{k,l}$ on $G^N$. In the proof, we will use the shorthand notation
$$
z
 \equiv 
\frac{1}{\sqrt{d_{\ul\lambda_1} d_{\ul\lambda_2}}}
 \,
\BF{\ul\lambda_1}{\lambda_1}{k_1,l_1}(\ul a)
 \,\, 
\BF{\ul\lambda_2}{\lambda_2}{k_2,l_2}(\ul a)
 \,,\quad
\Phi \equiv \Phi_{\ul\lambda_1\ul\lambda_2}
 \,,\quad
\Psi \equiv \Psi_{\ul\lambda_1\ul\lambda_2}
\,.
$$
Using \eqref{G-MF-1} and the fact that $\psi^1_{\ul\lambda_1\ul\lambda_2}$ is a $G^N$-representation isomorphism, we may rewrite 
$$
z
 =
\tr
 \left(
 \left(
\bigoplus_{\ul \lambda} \bigoplus_{i=1}^{m_{\ul\lambda_1,\ul\lambda_2}(\ul\lambda)}
 ~
\pi_{\ul\lambda}(\ul a) 
 \right)
 \circ 
\psi^1_{\ul\lambda_1\ul\lambda_2}
 \circ 
\left(A^{\ul\lambda_1,\lambda_1}_{k_1,l_1} \otimes A^{\ul\lambda_2,\lambda_2}_{k_2,l_2}\right)
 \circ 
(\psi^1_{\ul\lambda_1\ul\lambda_2})^{-1}
 \right)
\,.
$$
Since
$$
\bigoplus_{\ul \lambda} \bigoplus_{i=1}^{m_{\ul\lambda_1,\ul\lambda_2}(\ul\lambda)}
 ~
\pi_{\ul\lambda}(\ul a) 
 =
\sum_{\ul \lambda} \sum_{i=1}^{m_{\ul\lambda_1,\ul\lambda_2}(\ul\lambda)}
 ~
\mr i^{\ul\lambda_1,\ul\lambda_2}_{\ul\lambda,i}
 \circ 
\pi_{\ul\lambda}(\ul a) 
 \circ 
\pr^{\ul\lambda_1,\ul\lambda_2}_{\ul\lambda,i}
\,,
$$
this can be further rewritten as
$$
z
 =
\sum_{\ul \lambda} \sum_{i=1}^{m_{\ul\lambda_1,\ul\lambda_2}(\ul\lambda)}
\tr
 \left(
\pi_{\ul\lambda}(\ul a) 
 \circ 
\pr^{\ul\lambda_1,\ul\lambda_2}_{\ul\lambda,i}
 \circ 
\psi^1_{\ul\lambda_1\ul\lambda_2}
 \circ 
\left(A^{\ul\lambda_1,\lambda_1}_{k_1,l_1} \otimes A^{\ul\lambda_2,\lambda_2}_{k_2,l_2}\right)
 \circ 
(\psi^1_{\ul\lambda_1\ul\lambda_2})^{-1}
 \circ
\mr i^{\ul\lambda_1,\ul\lambda_2}_{\ul\lambda,i}
 \right)
.
$$
By \eqref{eq: rel21} and \eqref{eq: rel22},
 \ala{
\mr i^{\ul\lambda_1,\ul\lambda_2}_{\ul\lambda,i}
 & =
\sum_\lambda \sum_{k=1}^{m_{\ul\lambda}(\lambda)}
 ~
(\psi_{\ul\lambda_1\ul\lambda_2}^2)^{-1}
 \circ 
\mr{i}^{\ul\lambda_1,\ul\lambda_2}_{\ul\lambda,i,\lambda,k} 
 \circ 
\pr^{\ul\lambda}_{\lambda,k}
 \circ
\vp_{\ul\lambda}
\,,
\\
\pr^{\ul\lambda_1,\ul\lambda_2}_{\ul\lambda,i}
 & =
\sum_{\lambda} \sum_{k=1}^{m_{\ul\lambda}(\lambda)}
 ~
\vp_{\ul\lambda}^{-1} 
 \circ
\mr{i}^{\ul\lambda}_{\lambda,k}
 \circ 
\pr^{\ul\lambda_1,\ul\lambda_2}_{\ul\lambda,i,\lambda,k}
 \circ
\psi^2_{\ul\lambda_1\ul\lambda_2}
\,.
 }
Plugging this in, we obtain
 \ala{
z 
 = &
\sum_{\ul \lambda} 
\sum_{i=1}^{m_{\ul\lambda_1,\ul\lambda_2}(\ul\lambda)}
\sum_{\lambda,\lambda'} 
\sum_{k=1}^{m_{\ul\lambda}(\lambda)}
\sum_{l=1}^{m_{\ul\lambda}(\lambda')}
\tr
 \Big(
\pi_{\ul\lambda}(\ul a) 
 \circ 
\vp_{\ul\lambda}^{-1} 
 \circ
\mr{i}^{\ul\lambda}_{\lambda,k}
 \circ 
\pr^{\ul\lambda_1,\ul\lambda_2}_{\ul\lambda,i,\lambda,k}
 \circ
\left(\Psi \circ \Phi^{-1}\right)
\\
 & \hspace{1.5cm}
 \circ
\Phi
 \circ 
\left(A^{\ul\lambda_1,\lambda_1}_{k_1,l_1} \otimes A^{\ul\lambda_2,\lambda_2}_{k_2,l_2}\right)
 \circ 
\Phi^{-1}
 \circ 
\left(\Phi \circ \Psi^{-1}\right)
 \circ
\mr{i}^{\ul\lambda_1,\ul\lambda_2}_{\ul\lambda,i,\lambda,k} 
 \circ 
\pr^{\ul\lambda}_{\lambda,k}
 \circ
\vp_{\ul\lambda}
 \Big)
\,.
 }
Using \eqref{eq: prof-inj-z1-1} and \eqref{eq: prof-inj-z1-2}, we find 
$$
\Phi
 \circ
 \left(
A^{\ul\lambda_1,\lambda_1}_{k_1,l_1} \otimes A^{\ul\lambda_2,\lambda_2}_{k_2,l_2}
 \right)
 \circ
\Phi^{-1}
 = 
\frac{1}{\sqrt{d_{\lambda_1} d_{\lambda_2}}}
\sum_{\lambda''} \sum_{j=1}^{m_{(\lambda_1,\lambda_2)}(\lambda'')} 
 \!\!
\mr i^{\ul\lambda_1,\ul\lambda_2}_{\lambda_1\lambda_2,k_1k_2,\lambda'',j}
 \circ
\pr^{\ul\lambda_1,\ul\lambda_2}_{\lambda_1,\lambda_2,l_1,l_2,\lambda'',j}
\,.
$$
Together with \eqref{eq: UnitaryCoeff-1} and \eqref{eq: UnitaryCoeff-2} this yields, after taking the sums over $\lambda'$ and $\lambda''$, 
 \ala{
z 
 & =
\frac{1}{\sqrt{d_{\lambda_1} d_{\lambda_2}}}
\sum_{\ul \lambda} 
 ~
\sum_{i=1}^{m_{\ul\lambda_1,\ul\lambda_2}(\ul\lambda)}
 ~
\sum_{\lambda} 
 ~
\sum_{k,l=1}^{m_{\ul\lambda}(\lambda)}
 ~
\sum_{j=1}^{m_{(\lambda_1,\lambda_2)}(\lambda)}
U^{\ul\lambda_1,\ul\lambda_2;\ul\lambda,i,\lambda,k}_{\lambda_1,\lambda_2,k_1,k_2,\lambda,j}
 ~
\ol{U^{\ul\lambda_1,\ul\lambda_2;\ul\lambda,i,\lambda,l}_{\lambda_1,\lambda_2,l_1,l_2,\lambda,j}}
\\
 & \hspace{6.75cm}
\tr
 \Big(
\pi_{\ul\lambda}(\ul a) 
 \circ 
\vp_{\ul\lambda}^{-1} 
 \circ
\mr{i}^{\ul\lambda}_{\lambda,k}
 \circ 
\pr^{\ul\lambda}_{\lambda,k}
 \circ
\vp_{\ul\lambda}
 \Big)
\,.
 }
The assertion now follows from \eqref{A-T} and \eqref{ReprF-Hom}.
\ebw

\bbm\label{Bem-unitaries}

Note that the coefficients $U$ in Proposition \ref{MultLaw-Gen} depend on the unitary $G$-representation isomorphisms $\Phi_{\ul\lambda_1\ul\lambda_2}$ and $\Psi_{\ul\lambda_1\ul\lambda_2}$. In Subsection \ref{alg-SU2}, we will see that for $G = \SU(2)$, these isomorphisms are uniquely determined by the choice of a unitary $G$-representation isomorphism $\vp_{\ul\lambda}$ for every $\ul\lambda \in \widehat G^N$, and that the coefficients $U$ boil down to recoupling coefficients of angular momentum theory.
\qeb

\ebm


\section{The model for $G = \SU(2)$}
\label{model-SU2}



\subsection{The commutative algebra $\mc R$ for $G = \SU(2)$}
\label{alg-SU2}


As observed in the preceding section, to fix concrete basis functions $\BF{\ul\lambda}{\lambda}{ij}$, we have to fix the unitary $G$-representation isomorphisms $\vp_{\ul\lambda}$ entering their definition. As a consequence, we obtain concrete formulae for the unitary operators in the multiplication law of the above algebra, expressed in terms of $\SU(2)$-recoupling coefficients. This relates the algebra structure to the combinatorics of recoupling theory of angular momentum as provided in \cite{BL1,BL2,Louck,Yutsis}.

In the case of $G=\SU(2)$, the highest weights $\lambda$ of irreps correspond 1--1 to spins $j = 0 , \frac 1 2 , 1 , \frac 3 2 , \dots$. We will use the common notation $D^j$ for $\pi_j$. Thus, $(H_j,D^j)$ is the standard $\SU(2)$-irrep of spin $j$, spanned by the orthonormal ladder basis $\{\ket{j,m} : m = -j , -j+1 , \dots , j\}$ which is unique up to a phase. Accordingly, every  sequence $\ul\lambda$ of highest weights corresponds to a sequence $\ul j$ of spins. We write $D^{\ul j} \equiv \pi_{\ul j}$ for the corresponding irrep of $\SU(2)^N$ and $D_d^{\ul j} \equiv \pi^d_{\ul j}$ for the induced diagonal representation of $\SU(2)$. To fix the $G$-representation isomorphisms
\beq\label{G-vp1j}
\vp_{\ul j}
 : 
H_{\ul j}
 \to
\bigoplus_j \bigoplus_{i=1}^{m_{\ul j}(j)} H_j
\,,
\eeq
we choose the following reduction scheme for tensor products of $N$ irreps of $\SU(2)$. Given nonnegative half integers $s_1$, $s_2$, denote
$$
\langle s_1 , s_2 \rangle
 := 
\{|s_1-s_2| , |s_1-s_2|+1 , |s_1-s_2|+2 , \dots , s_1+s_2\}
$$
and recall that the representation space $H_{s_1} \otimes H_{s_2}$ decomposes into unique irreducible subspaces $(H_{s_1} \otimes H_{s_2})_s$ of spin $s \in \langle s_1 , s_2 \rangle$. We start with decomposing $H_{j^1}\otimes H_{j^2}$ into the unique irreducible subspaces $(H_{j^1}\otimes H_{j^2})_{l^2}$ with $l^2 \in \langle j^1 , j^2 \rangle$. Then, we decompose the invariant subspaces 
$$
(H_{j^1}\otimes H_{j^2})_{l^2} \otimes H_{j_3}
 \subset 
H_{j^1}\otimes H_{j^2}\otimes H_{j_3}
$$
into unique irreducible subspaces 
$$
((H_{j^1}\otimes H_{j^2})_{l^2} \otimes H_{j_3})_{l^3}
 \,,\quad
l^3 \in \langle l^2 , j^3 \rangle
\,.
$$
Iterating this, we end up with a decomposition of $H_{\ul j}$ into unique irreducible subspaces 
\beq\label{G-D-Hjl}
H_{\ul j , \ul l}
 :=
(
 \cdots
((H_{j^1}\otimes H_{j^2})_{l^2} \otimes H_{j_3})_{l^3} 
 \cdots \otimes 
H_{j^N})_{l^N}
\,,
\eeq
where $\ul l = (l^1 , \dots , l^N)$ is a sequence of nonnegative half integers satisfying $l^1 = j^1$ and $l^i \in \langle l^{i-1} , j^i\rangle$ for $i = 2 , 3 , \dots , N$. Let us denote the totality of such sequences by $R(\ul j)$. Moreover, denote
$$
\langle \ul j \rangle := \{j : \exists ~ \ul l \in R(\ul j) \text{ s.\ th.\ } j = l^N\}
 \,,\qquad
R(\ul j , j) = \{\ul l \in R(\ul j) : l^N = j\}
\,.
$$
Then, $m_{\ul j}(j) = |R(\ul j,j)|$ and hence $m_{\ul j}(j) \neq 0$ iff $j \in \langle \ul j \rangle$, and the isotypical component of type $j$ of $H_{\ul j}$ is given by the direct sum of the subspaces $H_{\ul j , \ul l}$ with $\ul l \in R(\ul j , j)$.

\bbm

Reduction schemes for $N$-fold tensor products of $\SU(2)$-irreps of spins $j^1 , \dots , j^N$ can be visualized by binary trees with $N$ terminal points $\circ$ labeled by $j^1 , \dots , j^N$ and representing the tensor factors, and with $N-1$ internal points $\bullet$ which have two incoming lines and, except for the last one, one outgoing line and which represent the intermediate reduction steps given by the irreducible subspaces in the tensor product of the incoming irreps. The last internal point represents the final irreducible subspace obtained by the reduction scheme. Every labeling of the internal points which is admissible in the sense that every internal label $l$ belongs to $\langle l^1 , l^2 \rangle$, where $l^1$ and $l^2$ label the starting points of the incoming lines, corresponds to a unique such final subspace. The binary tree of the reduction scheme used here is 
$$
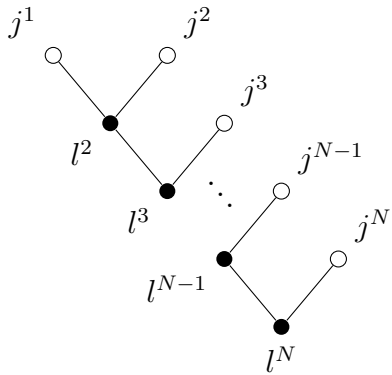
\begin{tikzpicture}[grow=up, level distance = 0.9 cm, sibling distance = 1.5cm, sloped]
\tikzstyle{bag} = [solid, circle, minimum width=6pt,fill, inner sep=0pt],
\tikzstyle{end} = [solid, circle, minimum width=6pt,inner sep=0pt, draw],
\tikzstyle{norm}= [edge from parent/.style ={black, thin, solid, draw},
		   every node/.style ={black, thin},
		   edge from parent path = {(\tikzparentnode) -- (\tikzchildnode)}],
\tikzstyle{spec}= [edge from parent/.style ={white, thin, draw},
		    every node/.style ={black,thin},
		    edge from parent path = {(\tikzparentnode) -- (\tikzchildnode)}]
 \node[ bag, label = below: {$l^N$}]{}
  child[norm]{node[end, label = above right: {$j^N$}]{}
	}
  child[norm]{
	node[bag, label = below left: {$l^{N-1}$}]{}
	child[norm]{node[end, label= above right:{$j^{N-1}$}]{}}
	child[spec]{node[bag, label =below left : {$l^3$}]{}
	      child[norm]{node[end, label=above right:{$j^3$}]{}}
	      child[norm]{node[bag, label = below left:{$l^2$}]{}
		    child[norm]{node[end, label = above right:{$j^2$}]{}}
		    child[norm]{node[end, label = above left:{$j^1$}]{}}
		  }
		  edge from parent
		  node[above= 0.5cm, left = -0.35 cm]{$\cdots$}
	      }
	}
;
\end{tikzpicture}
$$
and admissible internal labelings are given by the sequences $\ul l \in R(\ul j)$.
\qeb

\ebm

To define the isomorphism $\vp_{\ul j}$, we choose\footnote{Any other choice would yield the same basis vectors but multiplied by a phase which depends on $\ul l$ only.} ladder bases in the irreducible subspaces $H_{\ul j , \ul l}$. Denote their elements by $\ket{\ul j , \ul l , m}$, where $m = - l^N , -l^N+1 , \dots , l^N$. Then,
$$
\{\ket{\ul j , \ul l , m} : \ul l \in R(\ul j) , m = - l^N , -l^N+1 , \dots , l^N\}
$$
is an orthonormal basis in $H_{\ul j}$. For given $j \in \langle \ul j \rangle$, we can use the sequences $\ul l \in R(\ul j,j)$ to label the copies of $H_j$ in the direct sum decomposition of the target space of $\vp_{\ul j}$. As a consequence, the natural projections and injections related with this decomposition read $\pr^{\ul j}_{j,\ul l}$ and $\mr i^{\ul j}_{j,\ul l}$, respectively, the basis functions read $\BF{\ul j}{j}{\ul l,\ul l'}$ and the endomorphisms appearing in their definition read $A^{\ul j,j}_{\ul l,\ul l'}$. We define $\vp_{\ul j}$ by 
$$
\vp_{\ul j}(\ket{\ul j,\ul l,m}) := \mr i^{\ul j}_{j,\ul l}(\ket{j,m})
\,,
$$
where $\ket{j,m}$ denotes the elements of the orthonormal ladder basis in $H_j$. Using \eqref{A-T} and the relation $\pr^{\ul j}_{j,\ul l} \circ \mr i^{\ul j}_{j',\ul l'} = \delta_{\ul l,\ul l'} \id_{H_j}$, for $j \in \langle \ul j \rangle$, $\ul l , \ul l' \in R(\ul j,j)$ and $\ul l'' \in R(\ul j)$, we compute
$$
A^{\ul j,j}_{\ul l,\ul l'} (\ket{\ul j,\ul l'',m})
 =
\frac{1}{\sqrt{d_j}}
(\vp_{\ul j})^{-1} \circ \mr i^{\ul j}_{j,\ul l}
 \circ 
\pr^{\ul j}_{j,\ul l'} \circ \vp_{\ul j} (\ket{\ul j,\ul l'',m})
 =
\frac{\delta_{\ul l',\ul l''}}{\sqrt{d_j}} \, \ket{\ul j,\ul l,m}
\,.
$$
This implies 
\beq\label{Form-A}
A^{\ul j,j}_{\ul l,\ul l'}
 =
\frac{1}{\sqrt{d_j}} \, \sum_{m=-j}^j \ket{\ul j,\ul l,m} \bra{\ul j,\ul l',m}
 \,,\qquad
\ul l,\ul l' \in R(\ul j,j) \, ,
\eeq
and
\beq\label{Form-chi}
\BF{\ul j}{j}{\ul l,\ul l'}(\ul a)
 =
\sqrt{\frac{d_{\ul j}}{d_j}} \, \sum_{m=-j}^j 
\bra{\ul j,\ul l',m} D^{\ul j}(\ul a) \ket{\ul j,\ul l,m}
 \,,\qquad
\ul l,\ul l' \in R(\ul j,j)
\,.
\eeq
For later use, we express these functions in terms of the matrix entry functions $D^{j_i}_{m_i,m_i'}$, $i=1,\dots,N$. For spins $s_1 , s_2 , s$ and spin projections $m_1 , m_2 , m$, let 
$$
C^{s_1,s_2,s}_{m_1,m_2,m}
 :=
 \big\langle
\bra{s_1,m_1} \otimes \bra{s_2,m_2} \,
 \big|
{s_1,s_2;s,m}
 \big\rangle
$$
denote the Clebsch-Gordan coefficients. Here, $\ket{s_1,s_2;s,m}$ denote the elements of the ladder basis in the irreducible subspace of spin $s$ in $H_{s_1} \otimes H_{s_2}$ whenever $s \in \langle s_1,s_2\rangle$ and the zero vector otherwise.

\bsz\label{S-chi-D}

We have
$$
\BF{\ul j}{j}{\ul l,\ul l'}(\ul a)
 =
\sqrt{\frac{d_{\ul j}}{d_j}}
\sum_{m=-j}^j \sum_{\ul m} \sum_{\ul m'}
 \,
C(\ul j,\ul l,\ul m) \, C(\ul j,\ul l',\ul m')
 \,
D^{j_1}_{m_1',m_1}(a_1) \cdots D^{j_N}_{m_N',m_N}(a_N)
\,,
$$
where $\sum_{\ul m}$ means the sum over all sequences $\ul m = (m_1 , \dots , m_N)$ such that 
$$
m_i = - j^i , \dots , j^i \text{ for } i = 1 , \dots , N
 \,,\qquad
m_1 + \cdots + m_N = m
\,,
$$
and where
$$
C(\ul j,\ul l,\ul m)
 =
C^{j^1,j^2,l^2}_{m_1,m_2,m_1+m_2} 
C^{l^2,j^3,l^3}_{m_1+m_2,m_3,m_1+m_2+m_3}
 \cdots
C^{l^{N-1},j^N,l^N}_{m_1+\cdots+m_{N-1},m_N,m}
\,.
$$

\esz

\bbw

Using the tensor basis in $H_{\ul j}$, given by the vectors
$$
\ket{\ul j , \ul m} := \ket{j_1,m_1} \otimes \cdots \otimes \ket{j_N,m_N}
 \,,\qquad
m_i = -j_i , \dots , j_i \,,~ i = 1 , \dots , N
\,,
$$
formula \eqref{Form-chi} can be rewritten as
$$
\BF{\ul j}{j}{\ul l,\ul l'}
 =
\sqrt{\frac{d_{\ul j}}{d_j}} \, \sum_{m=-j}^j \,\sum_{\ul m} \sum_{\ul m'}
\braket{\ul j,\ul l',m}{\ul j,\ul m'} \braket{\ul j,\ul m}{\ul j,\ul l,m}
D^{j_1}_{m_1',m_1}(a_1) \cdots D^{j_1}_{m_N',m_N}(a_N)
\,.
$$
To compute the scalar products, we expand $\ket{\ul j,\ul l,m}$ wrt.\ $\ket{\ul j,\ul m}$. Denote $\undertilde j := (j^1 , \dots , j^{N-1})$ and $\undertilde l := (l^1 , \dots , l^{N-1})$ and consider the irreducible subspace $H_{\undertilde j,\undertilde l}$ of $H_{\undertilde j}$ with its ladder basis $\{\ket{\undertilde j,\undertilde l,m} : m = - l^{N-1} , \dots , l^{N-1}\}$. By construction, $H_{\ul j,\ul l}$ is the irreducible subspace of spin $l^N$ in $H_{\undertilde j,\undertilde l} \otimes H_{j^N}$ and $\ket{\ul j,\ul l,m}$ are the elements of the ladder basis in that subspace. Hence,
$$
\ket{\ul j,\ul l,m}
 = 
\sum_{n_N=-l^{N-1}}^{l^{N-1}} \sum_{m_N=-j^N}^{j^N}
C^{l^{N-1},j^N,l^N}_{n_N,m_N,m}
\ket{\undertilde j,\undertilde l,n_N} \otimes \ket{j^N,m_N}
\,.
$$
Iterating this argument, we find that the expansion of $\ket{\ul j,\ul l,m}$ is given by 
$$
\sum_{m_i,n_i}
 ~
C^{l^{N-1},j^N,l^N}_{n_N,m_N,m} 
C^{l^{N-2},j^{N-1},l^{N-1}}_{n_{N-1},m_{N-1},n_N}
 \cdots
C^{l^1,j^2,l^2}_{n_2,m_2,n_3}
 ~
\ket{j^1,n_2} \otimes \ket{j^2,m_2} \otimes \cdots \otimes \ket{j^N,m_N}
\,,
$$
where the sum runs over $n_i = - l^{i-1} , \dots , l^{i-1}$ and $m_i = -j^i , \dots , j^i$ for $i=2,\dots,N$. Putting $m_1=n_2$ and taking into account that the Clebsch-Gordan coefficients vanish unless the first two spin projections add up to the third one, we find that in the sum over $n_3 , \dots , n_N$, only the terms with 
$$
n_3 = m_1+m_2
 \,,\quad
n_4 = m_1+\dots+m_3
 \,,\quad \dots \,, \quad 
n_N = m_1 + \cdots + m_{N-1}
$$
survive. As a result, we obtain
$$
\ket{\ul j,\ul l,m}
 = 
\sum_{\ul m} ~ C(\ul j,\ul l,\ul m) ~ \ket{\ul j,\ul m}
\,.
$$
Plugging this into the above formula for $\BF{\ul j}{j}{\ul l,\ul l'}$ and taking into account that the Clebsch-Gordan coefficients are real, we obtain the assertion.
\ebw

Next, we compute the coefficients $U$ in the multiplication law for the basis functions given by Proposition \rref{MultLaw-Gen}. For that purpose, we have to determine the unitary $G$-representation isomorphisms $\Phi_{\ul j_1,\ul j_2}$ and $\Psi_{\ul j_1,\ul j_2}$ introduced in Subsection \rref{A-reprF}. 

First, consider $\Phi_{\ul j_1,\ul j_2}$. Recall that $\Phi_{\ul j_1,\ul j_2} = \phi_{\ul j_1,\ul j_2} \circ \big(\vp_{\ul j_1} \otimes \vp_{\ul j_2}\big)$. Here, $\phi_{\ul j_1,\ul j_2}$ is given by a unitary $G$-representation isomorphism
\beq\label{G-j1j2j}
H_{j_1} \otimes H_{j_2} \to \bigoplus_{j \in \langle j_1,j_2 \rangle} H_j
\eeq
for every pair $j_1,j_2$ with $j_1 \in \langle \ul j_1 \rangle$ and $j_2 \in \langle \ul j_2 \rangle$. Since the multiplicities are $1$ here, we may omit the corresponding index in our notation. Another consequence is that the isomorphism \eqref{G-j1j2j} is determined up to a phase on every $H_j$. We choose these phases in accordance with the standard choice of the Clebsch-Gordan coefficients, so that \eqref{G-j1j2j} is given by these coefficients. Then, $\Phi_{\ul j_1,\ul j_2}$ is uniquely determined by the choice of $\vp_{\ul j}$ for every $\ul j$ and hence by the choice of the reduction scheme for $N$-fold tensor products of $\SU(2)$-irreps. To write it down explicitly, we decompose $H_{\ul j_1} \otimes H_{\ul j_2}$ into irreducible subspaces according to the following reduction scheme: 
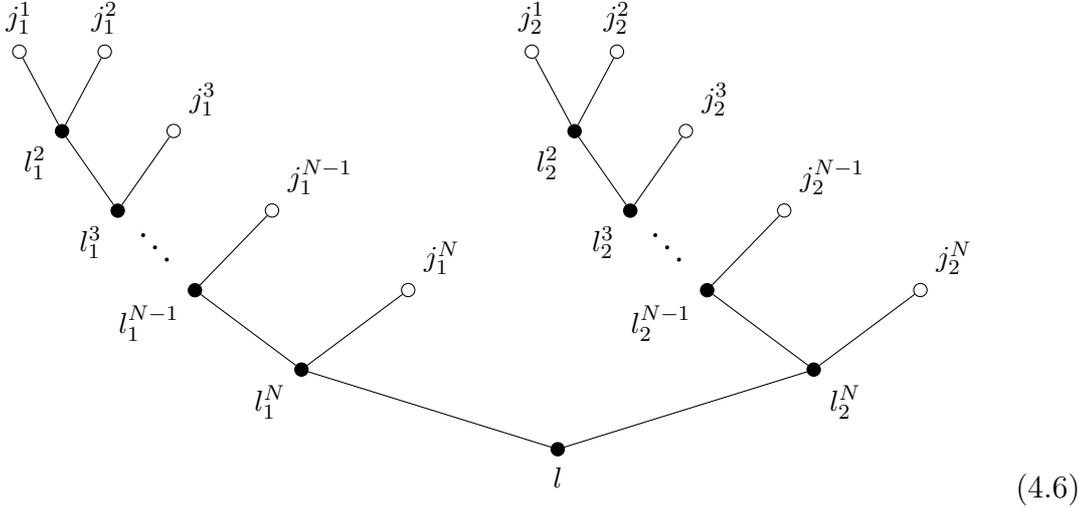
\beq\label{G-Baum-Phi}
\begin{tikzpicture}[grow= up,
  sibling distance =0.5cm,
  edge from parent/.style = draw, edge from parent path = {(\tikzparentnode) -- (\tikzchildnode)},
  sloped]
\tikzset{every node/.style = {font = \small}},
\tikzset{every internal node/.style ={solid,circle, minimum width = 5pt, fill, inner sep = 0pt, 						draw}},
\tikzset{every leaf node/.style = {solid, circle, minimum width = 5pt, inner sep = 0pt, draw}},
 \Tree
 [.\node[label = below:{$l$}]{};
  [.\node[label = below right: {$l_2^N$}]{};\node[label = above right:{$j_2^N$}]{};
    [.\node[label = below left:{$l_2^{N-1}$}]{}; \node[label = above right:{$j_2^{N-1}$}]{};
    \edge[white] node[black]{\large $\cdots$};
    [.\node[label = below left:{$l_2^{3}$}]{}; \node[label = above right:{$j_2^{3}$}]{};
    [.\node[label = below left:{$l_2^{2}$}]{}; \node[label = above:{$j_2^{2}$}]{};\node[label = above:{$j_2^1$}]{};]]]]
  [.\node[label = below left: {$l_1^N$}]{}; \node[label = above right:{$j_1^N$}]{};
    [.\node[label = below left:{$l_1^{N-1}$}]{}; \node[label = above right:{$j_1^{N-1}$}]{};
    \edge[white] node[black]{\large $\cdots$};
    [.\node[label = below left:{$l_1^{3}$}]{}; \node[label = above right:{$j_1^{3}$}]{};
    [.\node[label = below left:{$l_1^{2}$}]{}; \node[label = above:{$j_1^{2}$}]{};\node[label = above:{$j_1^1$}]{};]]]]]
\end{tikzpicture}
\eeq

This leads to irreducible subspaces labeled by $\ul l_1 \in R(\ul j_1)$, $\ul l_2 \in R(\ul j_2)$ and $l \in \langle l_1^N , l_2^N \rangle$. In each subspace, we choose an orthonormal ladder basis and denote its elements by $\ket{\ul j_1,\ul j_2 ; \ul l_1,\ul l_2 ; l,m}$, $m = -l , \dots , l$. Then,
$$
 \left\{
\ket{\ul j_1,\ul j_2 ; \ul l_1,\ul l_2 ; l,m}
 :
\ul l_1 \in R(\ul j_1) \,,~ \ul l_2 \in R(\ul j_2)
 \,,~ 
l \in \langle l_1^N , l_2^N \rangle
 \,,~ 
m = -l , \dots , l
 \right\}
$$
is an orthonormal basis in $H_{\ul j_1} \otimes H_{\ul j_2}$ and $\Phi_{\ul j_1,\ul j_2}$ is given by
$$
\Phi_{\ul j_1,\ul j_2}\left(\ket{\ul j_1,\ul j_2 ; \ul l_1,\ul l_2 ; l,m}\right)
 =
\mr i^{\ul j_1,\ul j_2}_{l_1^N,l_2^N,\ul l_1,\ul l_2,l}(\ket{l,m})
\,,
$$
where $\mr i^{\ul j_1,\ul j_2}_{l_1^N,l_2^N,\ul l_1,\ul l_2,j}$ denotes the natural injection associated with the decomposition \eqref{eq: decomp12} (here, by our specific choice of notation, the labels $l_1^N$ and $l_2^N$ are actually redundant). 

Now, consider the unitary $G$-representation isomorphism $\Psi_{\ul j_1,\ul j_2}$. Denote $\langle \ul j_1,\ul j_2 \rangle = \prod_{i=1}^N \langle j_1^i,j_2^i \rangle$.  Recall that $\Psi_{\ul j_1,\ul j_2} = \psi^2_{\ul j_1,\ul j_2} \circ \psi^1_{\ul j_1,\ul j_2}$, where 
$$
\psi^1_{\ul j_1,\ul j_2}
 : 
H_{\ul j_1} \otimes H_{\ul j_2}
 \to 
\bigoplus_{\ul j \in \langle \ul j_1,\ul j_2 \rangle} H_{\ul j}
$$
is a unitary $G^N$-representation isomorphism and $\psi^2_{\ul j_1,\ul j_2}$ acts on every summand $H_{\ul j}$ as $\vp_{\ul j}$. Since for every factor of $G^N$, $\psi^1_{\ul j_1,\ul j_2}$ boils down to an isomorphism of the type \eqref{G-j1j2j}, the multiplicities are $1$ as well and so we may omit the corresponding index in our notation. This also implies that $\psi^1_{\ul j_1,\ul j_2}$ is unique up to a phase for every factor of $G^N$ and the corresponding irreducible factor of $H_{\ul j}$. As before, we choose these phases so that $\psi^1_{\ul j_1,\ul j_2}$ is given by the appropriate Clebsch-Gordan coefficients. Then, $\Psi_{\ul j_1,\ul j_2}$, like $\Phi_{\ul j_1,\ul j_2}$, is uniquely determined by the choice of the reduction scheme for $N$-fold tensor products of $\SU(2)$-irreps. To write it down explicitly, we decompose $H_{\ul j_1} \otimes H_{\ul j_2}$ into irreducible subspaces according to the following reduction scheme: 
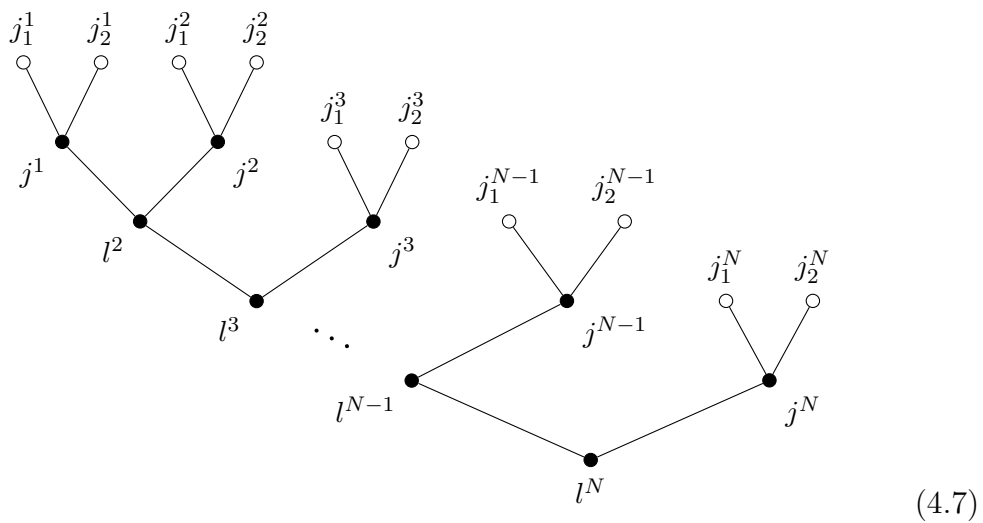
\beq\label{G-Baum-Psi}
\begin{tikzpicture}[grow= up,
  sibling distance = 0.4cm,
  edge from parent/.style = draw, edge from parent path = {(\tikzparentnode) -- (\tikzchildnode)},
  sloped]
\tikzset{every node/.style = {font = \small}},
\tikzset{every internal node/.style ={solid,circle, minimum width = 5pt, fill, inner sep = 0pt, 						draw}},
\tikzset{every leaf node/.style = {solid, circle, minimum width = 5pt, inner sep = 0pt, draw}},
\begin{scope}
\Tree
 [.\node[label = below:{$l^N$}]{};
    [.\node[label = below right:{$j^N$}]{}; \node[label = above:{$j^N_2$}]{};\node[label = above:{$j^N_1$}]{};] 
    [.\node[label = below left:{$l^{N-1}$}]{};[.\node[label = below right:{$j^{N-1}$}]{}; \node[label = above:{$j_2^{N-1}$}]{}; \node[label = above:{$j_1^{N-1}$}]{};]
      \edge[white] node[black]{\large$\cdots$};
      [.\node[label = below left:{$l^3$}]{};
      [.\node[label = below right:{$j^{3}$}]{}; \node[label = above:{$j_2^{3}$}]{}; \node[label = above:{$j_1^{3}$}]{};]
      [.\node[label = below left:{$l^2$}]{};
      [.\node[label = below right:{$j^{2}$}]{}; \node[label = above:{$j_2^{2}$}]{}; \node[label = above:{$j_1^{2}$}]{};]
      [.\node[label = below left:{$j^{1}$}]{}; \node[label = above:{$j_2^{1}$}]{}; \node[label = above:{$j_1^{1}$}]{};]]
      ]
      ]
    ]
\end{scope}
\end{tikzpicture}
\eeq
This leads to irreducible subspaces labeled by $\ul j \in \langle \ul j_1,\ul j_2 \rangle$ and $\ul l \in R(\ul j)$. In each subspace, we choose an orthonormal ladder basis and denote its elements by $\ket{\ul j_1,\ul j_2 ; \ul j,\ul l,m}$, $m = -l^N , \dots , l^N$. Then,
$$
 \left\{
\ket{\ul j_1,\ul j_2 ; \ul j,\ul l,m}
 ~:~
\ul j \in \langle \ul j_1,\ul j_2 \rangle
 \,,~ 
\ul l \in R(\ul j)
 \,,~ 
m = -l^N , \dots , l^N
 \right\}
$$
is an orthonormal basis in $H_{\ul j_1} \otimes H_{\ul j_2}$ and $\Psi_{\ul j_1,\ul j_2}$ is given by
$$
\Psi_{\ul j_1,\ul j_2}\left(\ket{\ul j_1,\ul j_2 ; \ul j,\ul l,m}\right)
 =
\mr i^{\ul j_1,\ul j_2}_{\ul j,l^N,\ul l}(\ket{l^N,m})
\,,
$$
where $\mr i^{\ul j_1,\ul j_2}_{\ul j,l^N,\ul l}$ denotes the natural injection associated with the decomposition \eqref{eq: decomp22} (where, by our specific choice of notation, the label $l^N$ is redundant).

\bsz\label{Gen-MultLaw}

In the case of $G = \SU(2)$, the multiplication law for the basis functions $\BFC{\ul j}{j}{\ul l,\ul l'}$ reads
 \al{\nonumber
\BFC{\ul j_1}{j_1}{\ul l_1,\ul l_1'} \cdot \BFC{\ul j_2}{j_2}{\ul l_2,\ul l_2'} 
 = &
\sqrt{\frac{d_{\ul{j_1}}d_{\ul{{j_2}}}}{d_{j_1}d_{j_2}}}
 ~
\sum_{\ul j \in \langle \ul j_1,\ul j_2 \rangle}
 ~
\sum_{j \in \langle j_1,j_2 \rangle}
	~
\sum_{\ul l,\ul l'\in R(\ul j, j)}
 ~
\sqrt{\frac{d_j}{d_{\ul j}}}
\\ \nonumber
& \hspace{3cm}
U_{\ul j_1,\ul j_2}(\ul j,\ul l;\ul l_1,\ul l_2)
 \,
U_{\ul j_1,\ul j_2}(\ul j,\ul l';\ul l_1',\ul l_2')
 \,
\BFC{\ul j}{j}{\ul l,\ul l'}
\,,
 }
where 
\beq\label{G-U-SU2}
U_{\ul j_1,\ul j_2}(\ul j,\ul l;\ul l_1,\ul l_2)
 =
\braket{\ul j_1,\ul j_2;\ul j,\ul l,m}{\ul j_1,\ul j_2;\ul l_1,\ul l_2;j,m}
\eeq
for every $\ul j \in \langle \ul j_1,\ul j_2 \rangle$, $j \in \langle j_1,j_2 \rangle$ and $\ul l \in R(\ul j, j)$, and for any admissible $m$.

\esz

The coefficients
$$
U_{\ul j_1,\ul j_2}(\ul j,\ul l;\ul l_1,\ul l_2)
 =
\braket{\ul j_1,\ul j_2;\ul j,\ul l,m}{\ul j_1,\ul j_2;\ul l_1,\ul l_2;j,m}
$$
are the recoupling coefficients for the reduction schemes \eqref{G-Baum-Phi} and \eqref{G-Baum-Psi}. Up to normalization, they are given by what is known as $3(2N-1)j$ symbols.\footnote{See Topic 12 in \cite{BL2} for details.}

\bbw

By Proposition \rref{MultLaw-Gen}, it suffices to compute the coefficients $U_{\ul j_1,\ul j_2}(\ul j,\ul l;\ul l_1,\ul l_2)$. According to \eqref{eq: UnitaryCoeff-1}, they are defined by 
$$
\pr^{\ul j_1,\ul j_2}_{\ul j,j,\ul l}
 \circ 
\Psi_{\ul j_1,\ul j_2} \circ \Phi_{\ul j_1,\ul j_2}^{-1}
 \circ 
\mr i^{\ul j_1,\ul j_2}_{j_1,j_2,\ul l_1,\ul l_2,j}
 =
U_{\ul j_1,\ul j_2}(\ul j,\ul l;\ul l_1,\ul l_2) \, \id_{H_j}
\,,
$$
where $\ul j \in \langle \ul j_1,\ul j_2 \rangle$, $j \in \langle j_1,j_2 \rangle$ and $\ul l \in R(\ul j, j)$. Evaluating the left hand side on a ladder basis vector $\ket{j,m}$ and plugging 
in a unit, we calculate
 \ala{
\pr^{\ul j_1,\ul j_2}_{\ul j,j,\ul l} 
 \circ 
\Psi_{\ul j_1,\ul j_2} & \circ \Phi_{\ul j_1,\ul j_2}^{-1}
 \circ 
\mr i^{\ul j_1,\ul j_2}_{j_1,j_2,\ul l_1,\ul l_2,j} \big(\ket{j,m}\big)
\\
 = \, &
\pr^{\ul j_1,\ul j_2}_{\ul j,j,\ul l} \circ \Psi_{\ul j_1,\ul j_2} 
\big(\ket{\ul j_1,\ul j_2;\ul l_1,\ul l_2;j,m}\big)
\\
 = \, &
\sum_{\ul j',\ul l'}
\braket{\ul j_1,\ul j_2;\ul j',\ul l',m}{\ul j_1,\ul j_2;\ul l_1,\ul l_2;j,m}
 \,
\pr^{\ul j_1,\ul j_2}_{\ul j,j,\ul l} \circ \Psi_{\ul j_1,\ul j_2} 
\big(\ket{\ul j_1,\ul j_2;\ul j',\ul l',m}\big)
\\
 = \, &
\sum_{\ul j',\ul l'}
\braket{\ul j_1,\ul j_2;\ul j',\ul l',m}{\ul j_1,\ul j_2;\ul l_1,\ul l_2;j,m}
 \,
\pr^{\ul j_1,\ul j_2}_{\ul j,j,\ul l} \circ \mr i^{\ul j_1,\ul j_2}_{\ul j',l'^N,\ul l'} 
\big(\ket{l'^N,m}\big)
\\
 = \, &
\braket{\ul j_1,\ul j_2;\ul j,\ul l,m}{\ul j_1,\ul j_2;\ul l_1,\ul l_2;j,m} \, \ket{j,m}
\,.
 }
This yields \eqref{G-U-SU2}. The multiplication law follows then by observing that the coefficients $U_{\ul j_1,\ul j_2}(\ul j,\ul l;\ul l_1,\ul l_2)$ are real. 
\ebw

The recoupling coefficients $U_{\ul j_1,\ul j_2}(\ul j,\ul l;\ul l_1,\ul l_2)$ can be expressed in terms of the recoupling coefficients for $N=2$, that is, for a tensor product of four $\SU(2)$-irreps as follows. Given four spins $j_1,j_2,j_4,j_5$, the tensor product $D^{j_1} \otimes D^{j_2} \otimes D^{j_4} \otimes D^{j_5}$ can be decomposed, on the one hand, into irreducible subspaces labeled by $j_3 , j_6 , j_9$ according to the following reduction scheme:
\beq\label{G-Baum-36}
\begin{tikzpicture}[grow= up,
  sibling distance = 0.4cm,
  edge from parent/.style = draw, edge from parent path = {(\tikzparentnode) -- (\tikzchildnode)},
  sloped]
\tikzset{every node/.style = {font = \small}},
\tikzset{every internal node/.style ={solid,circle, minimum width = 5pt, fill, inner sep = 0pt, 						draw}},
\tikzset{every leaf node/.style = {solid, circle, minimum width = 5pt, inner sep = 0pt, draw}},
\begin{scope}
\Tree
 [.\node[label = below:{$j_9$}]{};
    [.\node[label = below right:{$j_6$}]{}; \node[label = above:{$j_5$}]{};\node[label = above:{$j_4$}]{};] 
    [.\node[label = below left:{$j_3$}]{};\node[label = above:{$j_2$}]{};\node[label = above:{$j_1$}]{};] 
      ]
    ]
	]
\end{scope}
\end{tikzpicture}
\eeq
Let $\{\ket{(j_1,j_2),(j_4,j_5);j_3,j_6;j_9,m} : m = - j_9 , \dots , j_9\}$ be the ladder bases in these subspaces chosen in accordance with the definition of the Clebsch-Gordan coefficients\footnote{This is our notation specified to $N=2$. The common notation is $\ket{((j_1j_2)j_3,(j_4j_5)j_6)j_9,m}$.}. On the other hand, this tensor product can be decomposed into irreducible subspaces labeled by $j_7 , j_8 , j_9$ according to the following reduction scheme:
\beq\label{G-Baum-78}
\begin{tikzpicture}[grow= up,
  sibling distance = 0.4cm,
  edge from parent/.style = draw, edge from parent path = {(\tikzparentnode) -- (\tikzchildnode)},
  sloped]
\tikzset{every node/.style = {font = \small}},
\tikzset{every internal node/.style ={solid,circle, minimum width = 5pt, fill, inner sep = 0pt, 						draw}},
\tikzset{every leaf node/.style = {solid, circle, minimum width = 5pt, inner sep = 0pt, draw}},
\begin{scope}
\Tree
 [.\node[label = below:{$j_9$}]{};
    [.\node[label = below right:{$j_8$}]{}; \node[label = above:{$j_5$}]{};\node[label = above:{$j_2$}]{};] 
    [.\node[label = below left:{$j_7$}]{};\node[label = above:{$j_4$}]{};\node[label = above:{$j_1$}]{};] 
      ]
    ]
	]
\end{scope}
\end{tikzpicture}
\eeq
Let $\{\ket{(j_1,j_2),(j_4,j_5);(j_7,j_8),j_9,m} : m = - j_9 , \dots , j_9\}$ be the ladder bases in these subspaces, again chosen in accordance with the definition of the Clebsch-Gordan coefficients\footnote{In the common notation, $\ket{((j_1j_4)j_7,(j_2j_5)j_8)j_9,m}$.}. It is common to denote the recoupling coefficients between these two reduction schemes by
$$
\begin{pmatrix}
j_1 & j_2 & j_3
\\
j_4 & j_5 & j_6
\\
j_7 & j_8 & j_9
\end{pmatrix}
 :=
\braket{(j_1,j_2),(j_4,j_5);j_3,j_6;j_9,m}{(j_1,j_2),(j_4,j_5);(j_7,j_8),j_9,m}
$$
(the right hand side does not depend on $m$). These coefficients are related with Wigner's $9j$ symbols via
$$
\begin{pmatrix}
j_1 & j_2 & j_3
\\
j_4 & j_5 & j_6
\\
j_7 & j_8 & j_9
\end{pmatrix}
 =
\sqrt{(2j_3+1)(2j_6+1)(2j_7+1)(2j_8+1)}
\begin{Bmatrix}
j_1 & j_2 & j_3
\\
j_4 & j_5 & j_6
\\
j_7 & j_8 & j_9
\end{Bmatrix}.
$$

\ble\label{L-symbols}

The recoupling coefficients $U_{\ul j_1,\ul j_2}(\ul j,\ul l;\ul l_1,\ul l_2)$ are given by
$$
U_{\ul j_1,\ul j_2}(\ul j,\ul l;\ul l_1,\ul l_2)
 =
\prod_{i=2}^N
\begin{pmatrix}
l_1^{i-1} & l_2^{i-1} & l^{i-1}
\\
j_1^i & j_2^i & j^i
\\
l_1^i & l_2^i & l^i
\end{pmatrix}
,
$$
where $l_1^1 := j_1^1$, $l_2^1 := j_2^1$ and $l^1 := j^1$. 

\ele

\bbw

See Appendix \rref{Cor}.
\ebw

For general recursion formulae for $3nj$ symbols, see \cite{Yutsis}. Since there exist efficient calculators for $9j$ symbols, provided e.g.\ by the Python library SymPy \cite{sympy} or online by Anthony Stone's Wigner coefficient calculator\footnote{See http://www-stone.ch.cam.ac.uk/cgi-bin/wigner.cgi?symbol=9j}, Lemma \rref{L-symbols} provides an explicit knowledge of the multiplication law in the commutative algebra $\mc R$ for $\SU(2)^N$.


\subsection{The orbit type strata}
\label{SU2-OTStr}


Recall that for $G = \SU(2)$, we have $G_\CC = \SL(2,\CC)$, $\mf g = \su(2)$ and $\mf g_\CC = \sl(2,\CC)$. For convenience, we keep the notation $G$ and $G_\CC$. 
Let $Z$ denote the center of $G$. Clearly, this is also the center of $G_\CC$. Let $T \subset G$ denote the subgroup of diagonal matrices and let $\mf t$ be its Lie algebra. Clearly, $T$ is a maximal toral subgroup isomorphic to $\mr U(1)$.  

Let us briefly recall the orbit type strata $\pha_\tau$ of $\pha$ in terms of subsets $(G_\CC^N)_\tau$ of $G_\CC^N$. For details, see \cite{FuRS}.
First, one determines the orbit types of the lifted action of $G$ on $\ctg G^N$.  There are three of them and these can be labeled by $G$, $T$ and $Z$, where $Z$ is the principal orbit type. The corresponding orbit type subsets of $G^N \times \mf g^N$ are as follows.

\ben

\item[$(G)$]
An element $(\ul a , \ul A) \in G^N \times \mf g^N$ has orbit type $G$ iff 
$$
(\ul a , \ul A) \in Z^N \times \{0\}^N\,.
$$

\item[$(T)$] An element $(\ul a , \ul A) \in G^N \times \mf g^N$ has orbit type $T$ iff it is conjugate to an element of the subset 
$$
\left(T^N \times \mf t^N\right) \setminus \left(Z^N \times \{0\}^N\right)\,.
$$
Since conjugation by an element of $G$ commutes with taking commutators, for every element $(\ul a , \ul A)$ of orbit type $T$, the entries $(a_1, \ldots, a_N, A_1, \ldots , A_N)$ commute pairwise. Conversely, if for an element $(\ul a , \ul A)$ all its entries commute pairwise, then they are simultaneously diagonalizable and hence they belong to the orbit type $T$.

\item[$(Z)$] An element  $(\ul a , \ul A) \in G^N \times \mf g^N$ has orbit type $Z$ iff it does not have
orbit type $T$ or $G$, i.e., iff it is not conjugate to an element of $T^N \times
\mf t^N$, that is, iff not all entries of $(\ul a , \ul A)$ commute pairwise.  

\een

Next, one intersects the orbit type subsets with the momentum level set $\mu^{-1}(0)$, takes the quotient of $\mu^{-1}(0)$ with respect to the $G$-action, and passes to connected components. This yields the following.

\ben

\item[$(G)$] There exist $2^N$ orbit type strata of orbit type $G$, each of which consists of a single point representing the (trivial) orbit  of an element of $Z^N \times \{0\}^N$. Since such an element is of the form $(\nu_1 \II , \dots , \nu_N \II , 0, \dots , 0)$ for some sequence of signs $\ul\nu = (\nu_1 , \dots , \nu_N)$, we denote the corresponding stratum by $\pha_{\ul\nu}$. 

\item[$(T)$] Since $Z^N \times \{0\}^N$ consists of finitely many points and
$T^N \times \mf t^N$ has dimension at least $2$, the complement $(T^N \times \mf
t^N) \setminus (Z^N \times \{0\}^N)$ is connected. Since the subset of $\pha$ of
orbit type $T$ is the image of the subset 
$$
(T^N \times \mf t^N) \setminus (Z^N \times \{0\}^N) \subset \mu^{-1}(0)
$$
under the natural projection $\mu^{-1}(0) \to \pha$, it is connected,
too. Hence, it forms an orbit type stratum. We denote this stratum by $\pha_T$.

\item[$(Z)$] Since $\mf g^\ast$ has dimension $3$, the level set $\mu^{-1}(0)$ generically has dimension $2N \cdot 3 - 3 = 3(2N-1)$. On the other hand, since $T$ has dimension $1$ and the elements of $T^N \times \mf t^N$ have stabilizer $T$ under the action of $G$, the subset of $G^N \times \mf g^N$ of orbit type $T$ has dimension $2N \cdot 1 + (3-1) = 2(N+1)$. Hence, if the orbit type $Z$ occurs in $\pha$, i.e., if $N \geq 2$, then the subset of $\mu^{-1}(0)$ generated from $T^N \times \mf t^N$ by the action of $G$ has codimension
$$
3(2N-1) - 2(N+1) = 4N-5 \geq 3\,.
$$
Therefore, its complement is connected. Since the complement coincides with the subset of $\mu^{-1}(0)$ of orbit type $Z$, the subset of $\pha$ of this orbit type is connected. Hence, it forms an orbit type stratum. We denote this stratum by $\pha_Z$.

\een

One can visualize the set of strata and their partial ordering by a Hasse diagram, see \cite{FuRS}.
Finally, one transports the above results to $G_\CC^N$, that is, for each of the above strata, one finds the subset $(G_\CC^N)_\tau$ of $G_\CC^N$. It suffices to do this for every sequence of signs $\ul\nu = (\nu_1 , \dots , \nu_N)$ and for $T$. Let $T_\CC \subset G_\CC$ denote the subgroup of diagonal matrices. One obtains the following.

\btm\label{T-OT}

Let $\ul a \in G_\CC^N$. Then,

\ben

\sitem $\ul a \in (G_\CC^N)_{\ul \nu}$ iff $\ul a$ is orbit closure equivalent to $(\nu_1 \II , \dots , \nu_N \II)$,

\sitem $\ul a \in (G_\CC^N)_T$ iff $\ul a$ is orbit closure equivalent to an element of $T_\CC^N \setminus Z^N$.

\een

\etm

\subsection{Zero locus and radical ideal conditions}
\label{ZL-SU2}


In this subsection, for the strata $\tau$ found above, we recall from \cite{FuRS} the finite subsets $R_\tau$ of $\mc R$ having the corresponding orbit type subset $(G_\CC^N)_\tau$ as their zero locus and satisfying the radical ideal condition. 
Since $\tau = Z$ correponds to the principal stratum and hence $\mc H_Z = \mc H$, it suffices to discuss the secondary strata $\tau = \ul \nu$ and $\tau = T$.
For that purpose, we define the following $G$-invariant representative functions:
 \ala{
 &&
p^T_{rs}(\ul a) & := \tr\big([a_r,a_s]^2\big) \,,
 &
1 \leq r < s \leq N \, , \phantom{< t\,.}
 &&
\\
 &&
p^T_{rst}(\ul a) & := \tr\big([a_r,a_s]a_t\big) \,,
 &
1 \leq r < s < t \leq N\,.
 &&
 }

For the strata labeled by sequences of signs $\ul \nu$ one obtains the following.

\btm
\label{T-Rel-G}

The subset $(G_\CC^N)_{\ul \nu} \subset G_\CC^N$ is the set of common zeros of the $G_\CC$-invariant functions $p^T_{rs}$ with $1 \leq r < s \leq N$, $p^T_{rst}$ with $1 \leq r < s < t \leq N$ 
and 
$$
p^{\ul \nu}_r(\ul a) := \tr(a_r) - \nu_r 2
 \,,\qquad
r = 1 , \dots , N\,.
$$

\etm

\bbm
\label{Bem-OTnu}
 
Instead of using Theorem \rref{T-Rel-G}, one can construct the subspace $\mc H_{\ul \nu}$ associated with the stratum
$\pha_{\ul \nu}$ directly as follows.
Let $\{\psi_\alpha : \alpha \in A\}$ be an orthonormal basis of $\mc H$ which
contains a constant function $\psi_0$. Clearly, the basis provided by Proposition \rref{S-ReprF} is of that type. Since for a
continuous invariant function $\psi$, the condition to vanish on $(G_\CC^N)_{\ul
\nu}$ is equivalent to the condition $\psi(\nu_1 \II , \dots , \nu_N \II) = 0$,
the vanishing subspace $\mc V_{\ul \nu}$ of the stratum $\pha_{\ul \nu}$, given
by \eqref{G-V-GC}, is spanned by the elements 
$$
\psi_\alpha - \psi_\alpha(\nu_1 \II , \dots , \nu_N \II) \, 1
 \,,~~~~~~
\alpha \in A\,, ~ \alpha \neq 0\,,
$$
where $1$ denotes the constant function with value $1$. 
One proves that $\mc H_{\ul \nu}$ is spanned by the single element
$$
\psi_{\ul \nu} 
 =
\frac{1}{C_{\ul \nu}} \, 
\sum_{\beta \in A}
\ol{\psi_\beta(\nu_1 \II , \dots , \nu_N \II)} \, \psi_\beta\,,
$$
where $C_{\ul \nu}$ is a normalization constant. See Remark 5.4 in \cite{FuRS} for the details. 
\qeb

\ebm
By this remark, checking the radical ideal condition is relevant for the stratum $\pha_T$ only.

\btm\label{T-Rel-T}

The topological closure $\ol{(G_\CC^N)_T}$ is the set of common zeros of the $G$-invariant representative functions 
\beq\label{G-FnT}
p^T_{rs}
 \,,~
1 \leq r < s \leq N
 \,,\qquad
p^T_{rst}
 \,,~
1 \leq r < s < t \leq N
\,.
\eeq
The ideal in $\mc R$ generated  by these functions
is a radical ideal.

\etm
The proof of the radical ideal condition is the hard part of \cite{FuRS}.

For the construction of the subspace $\mc H_T$ associated with the stratum $T$, it will be convenient to express the functions $p^T_{rs}$ and $p^T_{rst}$ in 
terms of the basis functions $\BFC{\ul j}{j}{\ul l,\ul l'}$ introduced in Section \rref{alg-SU2}. It will turn out that the functions $p^T_{rs}$ are linear combinations 
of basis functions with $\ul j$ having entries $j_r$ at $r$, $j_s$ at $s$ and $0$ elsewhere. For such a sequence we write $\ul j = (j_rr,j_ss)$. The corresponding sequences 
$\ul l \in R(\ul j)$ have entries $l^1 = \cdots = l^{r-1} = 0$, $l^r = \cdots = l^{s-1} = j_r$ and $l^s = \cdots = l^N = j$, where $j \in \langle j_r,j_s \rangle$. Hence, 
for given $j$, there is only one sequence $\ul l$ in $R(\ul j,j)$, so that we may omit the labels $\ul l$ and $\ul l'$ in the notation. In a similar way, the functions $p^T_{rst}$ 
will turn out to be linear combinations of basis functions with $\ul j$ having entries $j_r$ at $r$, $j_s$ at $s$, $j_t$ at $t$ and $0$ elsewhere. For such a sequence we write 
$\ul j = (j_rr,j_ss,j_tt)$. Here, the sequences $\ul l \in R(\ul j)$ have entries $l^1 = \cdots = l^{r-1} = 0$, $l^r = \cdots = l^{s-1} = j_r$, $l^s = \cdots = l^{t-1} = l$ 
and $l^t = \cdots = l^N = j$, where $l \in \langle j_r,j_s \rangle$ and $j \in \langle l,j_s \rangle$. That is, they are labeled by a single intermediate spin $l$, so that 
in our notation we may replace the labels $\ul l$ and $\ul l'$ by $l$ and $l'$.

\ble
\label{L-Fuchs}

The functions $p^T_{rs}$ and $p^T_{rst}$ on $G^N_\mathbb{C}$ are given by
 \al{\label{rel1-inbasis}
p^T_{rs}
 & =
\BFC{(1r,0s)}{1}{}
 +
\BFC{(0r,1s)}{1}{}
 +
\BFC{(1r,1s)}{0}{}
 -
\frac{2}{\sqrt{3}} \BFC{(1r,1s)}{1}{} 
 -
3
\,,
\\ \label{rel2-inbasis}
p^T_{rst} 
 & = 
\frac{\sqrt{3}}{2}
\left(
\BFC{(\frac{1}{2}r,\frac{1}{2}s,\frac{1}{2}t)}{\frac{1}{2}}{0,1}
-
\BFC{(\frac{1}{2}r,\frac{1}{2}s,\frac{1}{2}t)}{\frac{1}{2}}{1,0}
\right)
\,.
 }

\ele

\bbw

According to the last statement of Corollary \rref{F-ReprF}, the expansion coefficients of $p^T_{rs}$ and $p^T_{rst}$ wrt.\ the basis $\{\BFC{\ul j}{j}{\ul l,\ul l'}\}$ in $\mc H$ coincide with the expansion coefficients of their restrictions to $G^N$ wrt.\ the basis $\{\BF{\ul j}{j}{\ul l,\ul l'}\}$ in $L^2(G^N)^G$. Hence, it suffices to determine the latter. By an abuse of notation, in what follows, $p^T_{rs}$ and $p^T_{rst}$ mean the restrictions to $G^N$.

First, consider $p^T_{rs}$. We have 
$$
p^T_{rs}=2 \tr\big((a_r a_s)^2\big) - 2 \tr(a_r^2 a_s^2)
\,.
$$
For the second term, we use $a = D^{\frac 1 2}(a)$ and $D^{\frac 1 2} \otimes D^{\frac 1 2} = D^0 \oplus D^1$ to calculate
\begin{align*}
\tr(a_r^2a_s^2)
 &=
\sum\limits_{m_i=\pm\frac12}
D^\frac{1}{2}_{m_1m_2}(a_r) D^\frac{1}{2}_{m_2m_3}(a_r)
D^\frac{1}{2}_{m_3m_4}(a_s) D^\frac{1}{2}_{m_4m_1}(a_s)
\\
 &=
\sum\limits_{m_i=\pm\frac12}
 \left\langle
\bra{\tfrac{1}{2},m_1} \otimes \bra{\tfrac{1}{2},m_2} 
 \left|
D^\frac{1}{2}(a_r)\otimes D^\frac{1}{2}(a_r)
 \right|
\ket{\tfrac{1}{2},m_2} \otimes \ket{\tfrac{1}{2},m_3}
 \right\rangle
\\
 & \hspace{2cm} \cdot
 \left\langle
\bra{\tfrac{1}{2},m_3} \otimes \bra{\tfrac{1}{2},m_4} 
 \left|
D^\frac{1}{2}(a_s)\otimes D^\frac{1}{2}(a_s)
 \right|
\ket{\tfrac{1}{2},m_4} \otimes \ket{\tfrac{1}{2},m_1}
 \right\rangle
\\
 &=
\sum\limits_{m_i=\pm\frac12}
 ~
\sum\limits_{j_r,j_s=0}^1 
 ~
\sum\limits_{n_r,n_r'=-j_r}^{j_r}
 ~
\sum\limits_{n_s,n_s'=-j_s}^{j_s}
\\
 & \hspace{2cm}
C^{\frac12 \frac12 j_r}_{m_1 m_2 n_r\phantom{'}} 
C^{\frac12 \frac12 j_r}_{m_2 m_3 n_r'} \,
C^{\frac12 \frac12 j_s}_{m_3 m_4 n_s\phantom{'}} 
C^{\frac12 \frac12 j_s}_{m_4 m_1 n_s'}
 \,
D^{j_r}_{n_r n_r'}(a_r) 
 \,
D^{j_s}_{n_s n_s'}(a_s)
\,.
\end{align*}
Hence, 
 \al{\label{eq: tr2}
\tr(a_r^2 a_s^2)
 & = 
\sum_{j_r,j_s =0}^1
 ~
\sum\limits_{n_r,n_r'=-j_r}^{j_r}
 ~
\sum\limits_{n_s,n_s'=-j_s}^{j_s}
 ~
R^{j_r j_s}_{n_r n_r' , n_s n_s'}
 \,
D^{j_r}_{n_r n_r'}(a_r) 
 \,
D^{j_s}_{n_s n_s'}(a_s) \, ,
\\ \nonumber
R^{j_r j_s}_{n_r n_r' , n_s , n_s'}
 & :=
\sum\limits_{m_i=\pm\frac12}
C^{\frac12 \frac12 j_r}_{m_1 m_2 n_r\phantom{'}} 
C^{\frac12 \frac12 j_r}_{m_2 m_3 n_r'} \,
C^{\frac12 \frac12 j_s}_{m_3 m_4 n_s\phantom{'}} 
C^{\frac12 \frac12 j_s}_{m_4 m_1 n_s'}
\,.
 }
A similar calculation for the first term of $p^T_{rs}$, using the relations
$$
C^{j_1j_2j}_{m_1m_2m}=(-1)^{j_1+j_2-j}C^{j_2j_1j}_{m_2m_1m}
 \,,\qquad
\sum_{m_1,m_2} C^{j_1 j_2 j}_{m_1 m_2 m} C^{j_1 j_2 j'}_{m_1 m_2 m'} = \delta_{jj'} \delta_{mm'}\,,
$$
yields
 \al{\label{eq: tr1}
\tr\big((a_r a_s)^2\big)
 & = 
\sum_{j_r,j_s =0}^1
 ~
\sum\limits_{n_r,n_r'=-j_r}^{j_r}
 ~
\sum\limits_{n_s,n_s'=-j_s}^{j_s}
 ~
S^{j_r j_s}_{n_r n_r' , n_s n_s'}
 \,
D^{j_r}_{n_r n_r'}(a_r) 
 \,
D^{j_s}_{n_s n_s'}(a_s)
\,,
\\ \nonumber
S^{j_r j_s}_{n_r n_r' , n_s n_s'}
 & :=
(-1)^{1-j_r} \delta_{j_r j_s} \delta_{n_r n_s'} \delta_{n_r' n_s}
\,.
 }
From \eqref{eq: tr2} and \eqref{eq: tr1} we conclude that $p^T_{rs}$ is a linear combination of the basis functions $\BF{(j_rr,j_ss)}{j}{}$ with $j_r , j_s = 0,1$ and $j \in \langle j_r , j_s\rangle$. To compute the expansion coefficients, we use Proposition \rref{S-chi-D} to write
\beq\label{G-chi-MEF-2}
\BF{(j_rr,j_ss)}{j}{}
 =
\sqrt{\frac{d_{j_r} d_{j_s}}{d_j}} 
 ~
\sum\limits_{m=-j}^j
 ~
\sum\limits_{n_r + n_s = m \atop n_r' + n_s' = m}
 ~
C^{j_r j_s j}_{n_r n_s m} C^{j_r j_s j}_{n_r' n_s' m}
 \,
D^{j_r}_{n_r n_r'}(a_r) 
 \,
D^{j_s}_{n_s n_s'}(a_s)
\eeq
and compute the scalar products $\braket{\BF{(j_rr,j_ss)}{j}{}}{p^T_{rs}}$ using the orthogonality relation 
\beq\label{G-OR}
\braket{D^j_{m_1 m_2}}{D^{j'}_{m_1' m_2'}}
 = 
\frac{1}{d_j} \, \delta_{jj'} \delta_{m_1 m_1'} \delta_{m_2 m_2'}
\,.
\eeq
This results in
$$
\braket{\BF{(j_rr,j_ss)}{j}{}}{p^T_{rs}}
 =
\frac{2}{\sqrt{d_{j_r} d_{j_s} d_j}}
\sum\limits_{m=-j}^j
 \,
\sum\limits_{n_r + n_s = m \atop n_r' + n_s' = m}
C^{j_r j_s j}_{n_r n_s m} C^{j_r j_s j}_{n_r' n_s' m}
 (
S^{j_r j_s}_{n_r n_r' , n_s , n_s'}-R^{j_r j_s}_{n_r n_r' , n_s , n_s'}
 )
\,.
$$
Computation of the right hand side for $j_r , j_s = 0,1$ and $j \in \langle j_r , j_s\rangle$ yields \eqref{rel1-inbasis}. For computations involving products of Clebsch-Gordan coefficients, one may use for example the Clebsch-Gordan coefficient function of Mathematica \cite{Mathematica} or the class sympy.physics.quantum.cg.CG provided by the Python library SymPy \cite{sympy}.
\bigskip

For $p^T_{rst}$ we proceed in an analogous way. Writing
$$
p^T_{rst}(\ul a) 
 =
\sum_{m_i = \pm \frac12}
 \left(
D^{\frac12}_{m_1 m_2}(a_r) D^{\frac12}_{m_2 m_3}(a_s) 
 -
D^{\frac12}_{m_1 m_2}(a_s) D^{\frac12}_{m_2 m_3}(a_r)
 \right)
D^{\frac12}_{m_3 m_1}(a_t)
\,,
$$
we see that $p^T_{rst}$ is a linear combination of the basis functions $\BF{(\frac12 r,\frac12 s,\frac12 t)}{j}{ll'}$ with admissible $j$, $l$ and $l'$, ie., with $j=\frac12$ and $l,l'=0,1$ or with $j=\frac32$ and $l=l'=1$ (five functions altogether). According to Proposition \rref{S-chi-D}, 
 \ala{
\BF{(\frac12 r,\frac12 s,\frac12 t)}{j}{ll'}
 & =
2 \, \sqrt{\frac{2}{d_j}} 
 ~
\sum\limits_{m=-j}^j
 ~
\sum\limits_{n_r + n_s + n_t = m \atop n_r' + n_s' + n_t' = m}
C^{\frac12 \frac12 l}_{n_r,n_s,n_r+n_s} C^{l \frac12 j}_{n_r+n_s,n_t,m}
\\
 & \hspace{2cm}
C^{\frac12 \frac12 l'}_{n_r',n_s',n_r'+n_s'} C^{l' \frac12 j}_{n_r'+n_s',n_t',m}
 \,
D^{\frac12}_{n_r n_r'}(a_r) 
 \,
D^{\frac12}_{n_s n_s'}(a_s)
 \,
D^{\frac12}_{n_t n_t'}(a_t)
\,.
 }
Using the orthogonality relation \eqref{G-OR}, we thus obtain
 \ala{
\braket{\BF{(\frac12 r,\frac12 s,\frac12 t)}{j}{ll'}}{p^T_{rst}}
 & =
\frac{1}{2 \sqrt{2 d_j}}
\sum\limits_{m=-j}^j
 ~
\sum\limits_{n_r + n_s + n_t = m \atop n_r' + n_s' + n_t' = m}
C^{\frac12 \frac12 l}_{n_r,n_s,n_r+n_s} C^{l \frac12 j}_{n_r+n_s,n_t,m}
\\
 & \hspace{2cm}
 \left(
C^{\frac12 \frac12 l'}_{n_t,n_r,n_r+n_t} C^{l' \frac12 j}_{n_r+n_t,n_s,m}
 -
C^{\frac12 \frac12 l'}_{n_s,n_t,n_s+n_t} C^{l' \frac12 j}_{n_s+n_t,n_r,m}
 \right)
\,.
 }
Now, \eqref{rel2-inbasis} follows by computing the right hand side for the values of $j$, $l$ and $l'$ given above.
\ebw


\subsection{The costratification}
\label{costrat}


According to Theorem \ref{T-Rel-T} and Corollary \ref{F-V}, the subspaces $\mc V_T$ and $\mc H_T$ associated with the stratum $\pha_T$ are given by 
\beq\label{G-V-SU2}
\mc V_T
 = 
\sum_{1\leq r < s \leq N} \im \big(\hat p^T_{rs} \big)
 + 
\sum_{1\leq r < s < t \leq N}\im \big(\hat p^T_{rst} \big) 
\eeq
and 
\beq\label{G-H-SU2}
\mc H_T
 = 
\bigcap_{1\leq r < s \leq N} \ker \big(\hat p^T_{rs}\big)^\dagger
 \cap 
\bigcap_{1\leq r < s < t \leq N} \ker\big(\hat p^T_{rst}\big)^\dagger
\,, 
\eeq
where the adjoint is taken with respect to the $L^2$-scalar product $\langle \cdot, \cdot \rangle$ defined by the measure $\nu_\hbar$ given by \eqref{measure-nu}. 

To derive $\mc V_T$ from \eqref{G-V-SU2} and $\mc H_T$ from \eqref{G-H-SU2} explicitly, we simplify the notation by collecting the data $\ul j$, $j$, $\ul l$ and $\ul l'$ labeling the basis functions in a multi-index
$$
I :=\left(\ul j ; j ; \ul l ; \ul l'\right).
$$
Let $\mc I$ denote the totality of all these multi-indices. According to Proposition \rref{Gen-MultLaw}, the structure constants of multiplication, defined by 
\beq
\label{StructConst}
\BFCi{I_1} \cdot \BFCi{I_2}
 = 
\sum_{I \in \mc I} ~ C^I_{I_1 I_2} \, \BFCi I
\,,
\eeq
are given by 
\beq\label{Gen-structC}
C^{I}_{I_1 I_2}
 ~=~
\sqrt{\frac{d_{\ul j_1} d_{\ul j_2} d_j}{d_{j_1} d_{j_2} d_{\ul j}}}
 ~
U_{\ul j_1,\ul j_2}(\ul j,\ul l;\ul l_1,\ul l_2)
 ~
U_{\ul j_1,\ul j_2}(\ul j,\ul l';\ul l_1',\ul l_2')
\,,
\eeq
where $I_i = \left(\ul j_i;j_i;\ul l_i;\ul l_i'\right)$ and $I = \left(\ul j;j;\ul l;\ul l'\right)$. According to \eqref{G-V-SU2}, the subspace $\mc V_T$ is spanned by the functions $p^T_{rs} \, \BFCi I$ with $1 \leq r < s \leq N$, $I \in \mc I$ and the functions $p^T_{rst} \, \BFCi I$ with $1 \leq r < s \leq N$, $I \in \mc I$. We expand 
\beq\label{Coeff}
p^T_{rs}
 = 
\sum_{K \in \mc I}  (p^T_{rs})^K \, \BFCi K
 \,,\qquad 
p^T_{rst} = \sum_{K \in \mc I} (p^T_{rst})^K \, \BFCi K
\,,
\eeq
where the coefficients $(p^T_{rs})^K$ and $(p^T_{rst})^K$ are given by Lemma \rref{L-Fuchs}. Then,
 \ala{
p^T_{rs} \, \BFCi I & = \sum_{J \in \mc I} A^J_I(r,s) \, \BFCi J 
 \,, &
A^J_I(r,s) & := \sum_{K \in \mc I} (p^T_{rs})^K C_{KI}^J
\,
\\
p^T_{rst} \, \BFCi I & = \sum_{J \in \mc I} B^J_I(r,s,t) \, \BFCi J
 \,, &
B^J_I(r,s) & := \sum_{K \in \mc I} (p^T_{rst})^K C_{KI}^J
\,.
 }
Thus, $\mc V_{T}$ is spanned by the functions
\beq\label{span-V-T}
\sum_{J \in \mc I} A^J_I(r,s) \, \BFCi J
 \,,~
r < s
 \,,\quad
\sum_{J \in \mc I} B^J_I(r,s,t) \, \BFCi J
 \,,~
r < s < t
 \,,\quad
I \in \mc I
\,.
\eeq

It remains to determine the coefficients $A^J_I(r,s)$ and $B^J_I(r,s,t)$. Recall the notation $(j_1r_1 , j_2r_2 , \dots)$ for a sequence of spins having entries $j_1$ at place $r_1$, $j_2$ at place $r_2$ etc., and $0$ elsewhere. In addition, we introduce the notation $\big(j_1|_{r_1}^{s_1} , j_2|_{r_2}^{s_2} , \dots\big)$ for a sequence of spins having entries $j_1$ at places $r_1 , \dots , s_1$, $j_2$ at places $r_2 , \dots , s_2$ etc., and $0$ elsewhere. From Lemma \rref{L-Fuchs}, we obtain

\btm\label{GenStrConst}

The vanishing subspace $\mc V_{T}$ is spanned by the functions \eqref{span-V-T}, with the  coefficients $A^J_I(r,s)$ and $B^J_I(r,s,t)$  given by
 \ala{
A^J_I(r,s)
 \,=~ & 
C^J_{I , \text{$\big($}(1r,0s);1;(1|_r^N);(1|_r^N)\text{$\big)$}}
 + 
C^J_{I , \text{$\big($}(0r,1s);1;(1|_s^N);(1|_s^N)\text{$\big)$}}
\\
 & 
 + 
C^J_{I , \text{$\big($}(1r,1s);0;(1|_r^{s-1});(1|_r^{s-1})\text{$\big)$}}
 + 
\frac{2}{\sqrt3} \, C^J_{I , \text{$\big($}(1r,1s);1;(1|_r^N);(1|_r^N)\text{$\big)$}}
 - 
3 \, \delta^J_I \, ,
\\
B^J_I(r,s,t)
 \,=~ &
\frac{\sqrt{3}}{2}
 \, 
C^J_{I , \text{$\big($}(\tinyfrac12r,\tinyfrac12s,\tinyfrac12t);\tinyfrac12;(\tinyfrac12|_r^{s-1},0_s^{t-1},\tinyfrac12|_t^N);(\tinyfrac12|_r^{s-1},1|_s^{t-1},\tinyfrac12|_t^N)\text{$\big)$}}
\\
 &
- \, \frac{\sqrt{3}}{2}
 \, 
C^J_{I , \text{$\big($}(\tinyfrac12r,\tinyfrac12s,\tinyfrac12t);\tinyfrac12;(\tinyfrac12|_r^{s-1},1_s^{t-1},\tinyfrac12|_t^N);(\tinyfrac12|_r^{s-1},0|_s^{t-1},\tinyfrac12|_t^N)\text{$\big)$}}.
 }
\etm

By taking the orthogonal complement, we obtain

\bfg\label{S-H-tau1}

The subspace $\mc H_{T}$ associated with the stratum $\pha_T$ consists of the vectors $\varphi = \varphi^J \, \BFCi J$ whose coefficients $\varphi^J$ are determined by the system of linear equations
$$
\sum_{J \in \mc I} \hat A^J_I(r,s) \, \varphi^J  = 0
 \,,~
r < s
 \,,\quad
\sum_{J \in \mc I} \hat B^J_I(r,s,t) \, \varphi^J  = 0
 \,,~
r < s < t
 \,,\quad
I \in \mc I
\,,
$$
where $\hat A^J_I(r,s) = A^J_I(r,s) \parallel \BFCi J \parallel^2$ and $\hat B^J_I(r,s,t) = B^J_I(r,s,t) \parallel \BFCi J \parallel^2$, with the norm $\parallel \BFCi J \parallel^2$ given by \eqref{norm}. 

\efg

\bbm\label{finite}

For given multi-indices $I_1 = (\ul j_1;j_1;\ul l_1;\ul l_1')$ and $I_2 = (\ul j_2;j_2;\ul l_2;\ul l_2')$, the range of $I_3 = (\ul j_3;j_3;\ul l_3;\ul l_3')$ for which the structure constant $C_{I_1 I_2}^{I_3}$ is nonzero is given by
$$
\ul{j_3} \in \langle \ul{j_1} , \ul{j_2} \rangle
 \,,\qquad 
j_3 \in \langle j_1 , j_2 \rangle
 \,,\qquad 
\ul l_3 , \ul l_3' \in R(\ul{j_3},j_3)
\,.
$$
In particular, the range is finite. Hence, the sums in Theorem \ref{GenStrConst} are finite. Furthermore, for fixed $I_1$ and $I_3$ the range of $I_2$ is finite, too, because $\ul{j_2}$ and $j_2$ are bounded by
$$
|j_1^{\,i}-j_2^{\,i}| \leq j_3^i \,,~ i = 1 , \dots , N
 \,,\qquad
|j_1-j_2|\leq j_3
$$
and the range of the sequences $\ul l_2$, $\ul l_2'$ of intermediate spins is given by $R(\ul{j_2},j_2)$. Since $C_{I_1 I_2}^{I_3} = C_{I_2 I_1}^{I_3}$, for fixed $I_2$ and $I_3$, the range of $I_1$ is bounded as well.

In contrast, the sums in Corollary \ref{S-H-tau1} are not finite. To find the coefficients $\vp^J$, one has to rewrite the defining equations into recurrence relations and 
to use the asymptotic behaviour of the norms. This will be discussed elsewhere. 
\qeb

\ebm

To summarize, the costratification for $G = \SU(2)$ consists of the Hilbert subspaces
$$
{\mc H}_{\ul \nu} \, ,\quad {\mc H}_T\, , \quad  {\mc H}_Z = \mc H \, ,
$$
together with their orthogonal projectors. Here, ${\mc H}_{\ul \nu}$ is given by Remark \ref{Bem-OTnu} and ${\mc H}_T$ is given by Theorem \ref{S-H-tau1}. By Lemma 7 of \cite{Fuchs}, the orthogonal projector onto the vanishing subspaces of the point strata are given by 
$$
{\mathbbm P}_{\ul \nu} (f) := f - f(\nu_1 {\mathbbm 1}, \ldots, \nu_N {\mathbbm 1}) \, 1
\,,
$$
where $1$ denotes the constant function on $G^N_\CC$ with value one. Thus, the projector onto $\mc H_{\ul \nu}$ is $\id_{\mc H} - {\mathbbm P}_{\ul \nu}$. There are various approaches to the 
construction of the orthogonal projector for the $T$-stratum. One of them consists in applying the Schmidt orthogonalization procedure to the family \eqref{span-V-T}. This will be studied 
in future work.


\subsection{The case $N=2$}
\label{N-2}


To illustrate the general result, let us discuss the case $N = 2$. Here, for given $\ul j = (j^1,j^2)$ and $j \in \langle \ul j \rangle \equiv \langle j^1,j^2 \rangle$, the set $R(\ul j,j)$ consists of the single sequence $\ul l = (j)$. Hence, the labels $\ul l$ and $\ul l'$ are redundant and the basis functions may be denoted by $\BFC{(j^1,j^2)}{j}{}$. Moreover, the isotypical components  of $H_{\ul j} = H_{j^1} \otimes H_{j^2}$ are irreducible and the endomorphisms $A^{\ul j,j}_{\ul l,\ul l'}$ boil down to orthogonal projectors $A^{(j^1,j^2),j}$. Thus, the basis functions are given by 
$$
\BF{(j^1,j^2)}{j}{}(a_1,a_2)
 = 
\sqrt{\frac{d_{j^1} d_{j^2}}{d_j}}
 \,
\tr\Big(A^{(j^1,j^2),j} \circ \big(D^{j^1}(a_1) \otimes D^{j^2}(a_2)\big)\Big)
\,.
$$
By \eqref{norm},
\beq
\label{norm-N2}
\|\BFC{(j^1,j^2)}{j}{}\|^2
 = 
(\hbar \pi)^3 \mr e^{\hbar\big((2j^1+1)^2 + (2j^2+1)^2\big)}
\,.
\eeq
According to Lemma \rref{L-symbols}, the recoupling coefficients $U$ are given by 
$$
U_{\ul j_1,\ul j_2}(\ul j, j;j_1, j_2)
 =
\begin{pmatrix} 
j_1^1 & j_2^1 & j^1 \\  j_1^2 & j_2^2 & j^2 \\ j_1 & j_2 & j 
\end{pmatrix}
.
$$
Hence, by Proposition \rref{Gen-MultLaw}, the multiplication law reads
\beq\label{Mult-law}
\BFC{\ul j_1}{j_1}{} \cdot \BFC{\ul j_2}{j_2}{}
 ~=~
\sum_{j^1 \in \langle j_1^1,j_2^1 \rangle \atop j^2 \in \langle j_1^2,j_2^2 \rangle}
 ~
\sum_{j \in \langle j^1,j^2 \rangle}
 ~
\bbma
j_1^1 & j_2^1 & j^1 \\  j_1^2 & j_2^2 & j^2 \\ j_1 & j_2 & j
\ebma
\BFC{\ul j}{j}{}
\,,
\eeq
where
$$
\bbma
j_1^1 & j_2^1 & j^1 \\  j_1^2 & j_2^2 & j^2 \\ j_1 & j_2 & j
\ebma
 =
 \sqrt{\frac{
d_{j_1^1} d_{j_1^2} d_{j_2^1} d_{j_2^2} d_{j}
 }{
d_{j_1} d_{j_2} d_{j^1} d_{j^2}
 }}
\begin{pmatrix} 
j_1^1 & j_2^1 & j^1 \\  j_1^2 & j_2^2 & j^2 \\ j_1 & j_2 & j
\end{pmatrix}^2
$$
is the structure constant $C^{(\ul j;j)}_{(\ul j_1;j_1),(\ul j_2;j_2)}$. In terms of Wigner's $9j$ symbols,
$$
\bbma j_1 & j_2 & j_3 \\ j_4 & j_5 & j_6 \\ j_7 & j_8 & j_9 \ebma
 =
\sqrt{\prod_{i=1}^9 d_{j_i}} 
 \,
\begin{Bmatrix}
j_1 & j_2 & j_3 \\ j_4 & j_5 & j_6 \\ j_7 & j_8 & j_9
\end{Bmatrix}^2
\,.
$$
By the analysis of Subsection \ref{SU2-OTStr}, we have the following orbit types and strata.

\ben

\item[$(G)$] The stabilizer is $G=\SU(2)$. The corresponding subset of $\pha$ 
consists of (the trivial orbits of) the points 
$$
\big((a,b),(A,B)\big) = \big(\pm\II,\pm\II),(0,0)\big)\,.
$$
Hence, this subset decomposes into the four strata $\pha_{\pm\pm}$. 

\item[$(T)$] The stabilizer is a torus. The corresponding stratum $\pha_T$ consists of the orbits of the points $\big((a_1,a_2),(A_1,A_2)\big) \neq \big((\pm\II,\pm\II,(0,0)\big)$ for which $a_1,a_2,A_1,A_2$ commute pairwise.

\item[$(Z)$] The stabilizer is the center of $\SU(2)$. The corresponding stratum $\pha_0$ consists of the orbits of the points $\big((a_1,a_2),(A_1,A_2)\big)$ for which $a_1,a_2,A_1,A_2$ do not commute pairwise.

\een

To these strata, there correspond the following closed subspaces of $\mc H$.

\ben

\item[$(Z)$] As in the general case, the subspace $\mc H_0$ associated with the principal stratum $\pha_0$ coincides with $\mc H$. 

\item[$(G)$] The subspaces $\mc H_{\pm\pm}$ associated with the strata $\pha_{\pm\pm}$ can be constructed as outlined in Remark \ref{Bem-OTnu}. Since $\pha_{\pm\pm}$ corresponds to $(\pm\II,\pm\II) \in G_\CC \times G_\CC$, the subspace $\mc V_{\pm\pm}$ is spanned by the functions 
$$
\BFC{(j^1,j^2)}{j}{} - \BFC{(j^1,j^2)}{j}{}(\pm\II,\pm\II) \, 1
$$
and the subspace $\mc H_{\pm\pm}$ is spanned by the single vector  
$$
\psi_{\pm\pm}
 = 
\frac{1}{N_{\pm\pm}}
 \,
\sum_{j^1,j^2} \sum_{j \in \langle j^1,j^2 \rangle}
 \,
\ol{\BFC{(j^1,j^2)}{j}{}(\pm\II,\pm\II)} \, \BFC{(j^1,j^2)}{j}{}
\,.
$$

\item[$(T)$] The subspace $\mc H_T$ associated with the stratum $\pha_T$ is defined by the single function
$$
p^T(a_1,a_2) = \tr\big([a_1,a_2]^2\big)
\,.
$$
By Lemma \ref{L-Fuchs}, 
\beq
\label{rel1-inbasis-2}
p^T
 = 
\BFC{(1,0)}{1}{}
 + 
\BFC{(0,1)}{1}{}
 + 
\BFC{(1,1)}{0}{}
 - 
\frac{2}{\sqrt 3} \, \BFC{(1,1)}{1}{}
 -
3
\,.
\eeq
As a consequence, the vanishing subspace $\mc V_T$ is spanned by the vectors 
$$
\sum_{J \in\mc I} A^J_I \BFCi J
 \,,\quad
I \in \mc I
\,,
$$
where according to Proposition \rref{GenStrConst}, the coefficients $A^J_I$ are given by 
$$
A^J_I
 =  
\bbma 1 & i^1 & j^1 \\ 0 & i^2 & j^2 \\ 1 & i & j \ebma
 + 
\bbma 0 & i^1 & j^1 \\ 1 & i^2 & j^2 \\ 1 & i & j \ebma
 + 
\bbma 1 & i^1 & j^1 \\ 1 & i^2 & j^2 \\ 0 & i & j \ebma
 + 
\frac{2}{\sqrt{3}} \bbma 1 & i^1 & j^1 \\ 1 & i^2 & j^2 \\ 1 & i & j \ebma
 - 
3 \delta^J_I
$$
with $L = (\ul i;i)$ and $J = (\ul j;j)$. Finally, Corollary \ref{S-H-tau1} implies that $\mc H_T$ consists of the functions $\varphi = \varphi^J \BFCi J$ whose coefficients $\varphi^J$ are determined by the system of linear equations
$$
\sum_{J \in \mc I} \hat A^J_I \varphi^J = 0 
 \,,\quad
I \in \mc I
\,,
$$
where $\hat A^J_I = A^J_I \parallel \BFCi J \parallel^2$ and $\parallel \BFCi J \parallel^2$ is given by \eqref{norm-N2}. 

\een


\subsection{The eigenvalue problem for the Hamiltonian}
\label{Hamilton}


Recall the classical Hamiltonian
$$
H(a,E)
 = 
\frac{g^2}{2 \delta} \sum_{\ell \in \Lambda^1} \|E(\ell)\|^2
 -
\frac{1}{g^2 \delta} \sum_{p \in \Lambda^2} \left(\tr a(p) + \ol{\tr a(p)}\right)
 \,,
$$
given by \eqref{Hamiltonian-C}. Here, $a(p)= a(\ell_1)a(\ell_2)a(\ell_3)a(\ell_4)$,  where the links $\ell_1 , \dots , \ell_4$, in this order, form the boundary of $p$ and are endowed with the boundary orientation. The quantum Hamiltonian, obtained via canonical quantization in the tree gauge, is called the Kogut-Susskind Hamiltonian (more precisely, its pure gauge part):
\beq
\label{Hamiltonian}
H = \frac{g^2}{2\delta} \, \mf C - \frac{1}{g^2 \delta} \, \mf W
\,.
\eeq
Here,
$$
\mf C := \sum_{\ell \in \Lambda^1} E_{ij}(\ell) E_{ji}(\ell)
$$
is the Casimir operator (negative of the group Laplacian) of $\SU(2)^N$ and 
$$
\mf W := \sum_{p \in \Lambda^2} ( W (p) + W(p)^*)
\,,
$$
where $W(p)$ is the quantum counterpart (multiplication operator on $\mc H$) of $\tr a(p)$, called the Wilson loop operator. For details, see \cite{qcd3}, \cite{GR2}, \cite{KS}. Recall that the representative functions of spin $j$ on $SU(2)$ are eigenfunctions of the Casimir operator of $\SU(2)$ corresponding to the eigenvalue\footnote{See \cite{Helgason}, \cite{Fegan}}
$$
\epsilon_j = 4 j(j+1)
\,.
$$
It follows that the invariant representative functions $\BF{\ul j}{j}{\ul l,\ul l'}$ are eigenfunctions of ${\mathfrak C}$ corresponding to the eigenvalues
\beq
\label{Eigenvalue-Casimir}
\epsilon_{\ul j} = \epsilon_{j^1} + \cdots + \epsilon_{j^N}
\,. 
\eeq
Let us analyze $\mf W$. For that purpose, for our regular cubic lattice, we define a standard tree as follows. By a line we mean a maximal straight line consisting of lattice links. First, choose a lattice site $x_0$ and a line $L_1$ through $x_0$. Next, choose a second line $L_2$ through $x_0$ perpendicular to $L_1$ and add all lines parallel to $L_2$ in the plane spanned by $L_1$ and $L_2$. Finally, add all lines perpendicular to that plane. Let  $B$ be such a standard tree. Since 
$a(\ell) = \mathbbm{1}$ for every $\ell \in B^1$, we can decompose $\mathfrak{W}$ into three sums.\footnote{Note that, for the standard tree, no plaquettes 
having 3 off-tree links occur.} It is easy to check that there exists an orientation and a numbering of the off-tree links such that for every plaquette with four off-tree links (all of these plaquettes are parallel to the plane spanned by the lines $L_1$ and $L_2$), the boundary links are numbered and oriented consistently, meaning that for one of the two possible orientations of the plaquette, they carry the induced boundary orientation, and that their numbers increase in that direction. Then, 
\begin{align*}
\mathfrak{W}
 =~ & 
\sum_{\{p:\,p \cap B =\emptyset\}} 
\tr(a_{r_p} a_{s_p} a_{t_p} a_{u_p})
 +
\overline{\tr(a_{r_p} a_{s_p} a_{t_p} a_{u_p})}
\\
 &
+ \sum_{\{p:\,|p \cap B|=2\}}\tr(a_{r_p} a_{s_p}) + \overline{\tr(a_{r_p} a_{s_p})}
\\
 &
+ \sum_{\{p:\,|p \cap B|=3\}}\tr(a_{r_p})+\overline{\tr(a_{r_p})}.
\end{align*}
To find the matrix elements of $H$ with respect to the basis functions $\{\BF{\ul j}{j}{\ul l,\ul l'}$, we have to find the corresponding expansion of $\mf W$. The sequences $\ul j$ occuring here will have at most four nonzero entries, so we can use the notation introduced in Section \rref{ZL-SU2}, given by writing $\ul j = (j_1r_1,\dots,j_kr_k)$ if $\ul j$ has entries 0 except for $j_i$ at the places $r_i$, $i=1,\dots,k$. The function $T_r(\ul a) = \tr(a_r)$ coincides with the basis function $\BF{\ul j}{j}{\ul l,\ul l'}$ with $\ul j = (\frac12 r)$. Omitting the irrelevant indices $j , \ul l , \ul l'$, we thus have 
$$
T_r = \BFi{(\frac12 r)}
\,.
$$
The function $T_{rs}(\ul a) := \tr(a_r a_s)$ is a linear combination of the basis functions $\BF{\ul j}{j}{\ul l \ul l'}$ with $\ul j = (\frac12r,\frac12s)$. As in Section \rref{ZL-SU2}, we may omit the irrelevant labels $\ul l$, $\ul l'$. Using \eqref{G-chi-MEF-2}, the orthogonality relation for the matrix entry functions given by \eqref{G-OR} and the normalization condition $\sum_{m_1,m_2=\pm\frac12} (C^{\frac12 \frac12 j}_{m_1 m_2 m})^2 = 1$, we obtain
$$
\braket{\BF{(\frac12r,\frac12s)}{j}{}}{T_{rs}}
 =
(-1)^{1-j} \, \frac{\sqrt{d_j}}{2}\,.
$$
Thus,
$$
T_{rs}
 = 
\frac{\sqrt 3}{2} \, \BF{(\frac12 r,\frac12 s)}{1}{}
 -
\frac 1 2 \, \BF{(\frac12 r,\frac12 s)}{0}{}
\,.
$$
Finally, the function $T_{rstu}(\ul a) := \tr(a_r a_s a_t a_u)$ is a linear combination of the basis functions $\BF{\ul j}{j}{\ul l \ul l'}$ with $\ul j = (\frac12r,\frac12s,\frac12t,\frac12u)$. Here, the sequences $\ul l \in R(\ul j,j)$ have entries $l^1 = \cdots = l^{r-1} = 0$, $l^r = \cdots = l^{s-1} = \frac12$, $l^s = \cdots = l^{t-1} = l$, $l^t = \cdots = l^{u-1} = k$ and $l^u = \cdots = l^N = j$, where $l =0,1$ and $k \in \langle l , \frac12 \rangle$ so that $j \in \langle k , \frac12 \rangle$. That is, they are labeled by two intermediate spins $l,k$, so that in our notation we may replace the labels $\ul l$ and $\ul l'$ by $(l,k)$ and $(l',k')$, respectively. Expressing these basis functions in terms of matrix entry functions according to Proposition \rref{S-chi-D} and using once again the orthogonality relation \eqref{G-OR}, we obtain
 \ala{
\braket{\BF{(\frac12r,\dots,\frac12u)}{j}{(l,k),(l',k')}}{T_{rstu}}
 & =
\frac{1}{4 \sqrt{d_j}}
 \,
\sum_{m=-j}^j ~ \sum_{m_r+\cdots+m_u=m} ~ 
\\
 & \hspace{1cm}
C^{\frac12 \frac12 l}_{m_r,m_s,m_r+m_s}
C^{l \frac12 k}_{m_r+m_s,m_t,m_r+m_s+m_t}
C^{k \frac12 j}_{m_r+m_s+m_t,m_u,m}
\\
 & \hspace{1cm}
C^{\frac12 \frac12 l'}_{m_u,m_r,m_r+m_u}
C^{l' \frac12 k'}_{m_r+m_u,m_s,m_r+m_s+m_u}
C^{k' \frac12 j}_{m_r+m_s+m_u,m_t,m}
\,.
 }
Evaluation yields
\begin{align*}
 T_{rstu}
 \,=~ & \textstyle
\frac 1 8 \, \BF{(\frac12r,\cdots,\frac12u)}{0}{(0\frac12)(0\frac12)}
 -
\frac{\sqrt{3}}{8} \, \BF{(\frac12r,\dots,\frac12u)}{0}{(1\frac12)(0\frac12)}
 -
\frac{\sqrt{3}}{8} \, \BF{(\frac12r,\dots,\frac12u)}{0}{(0\frac12)(1\frac12)}
\\
 & \textstyle
 - 
\frac 1 8 \BF{(\frac12r,\dots,\frac12u)}{0}{(1\frac12)(1\frac12)}
 -
\frac{\sqrt{3}}{8} \BF{(\frac12r,\dots,\frac12u)}{1}{(0\frac12)(0\frac12)}
 -
\frac 1 8 \BF{(\frac12r,\dots,\frac12u)}{1}{(1\frac12)(0\frac12)}
\\
 & \textstyle
 -
\frac{1}{\sqrt{8}} \BF{(\frac12r,\dots,\frac12u)}{1}{(1\frac32)(0\frac12)}
 +
\frac 3 8 \BF{(\frac12r,\dots,\frac12u)}{1}{(0\frac12)(1\frac12)}
 -
\frac{1}{8\sqrt{3}} \BF{(\frac12r,\dots,\frac12u)}{1}{(1\frac12)(1\frac12)}
\\
 & \textstyle
 -
\frac{1}{2\sqrt{6}} \BF{(\frac12r,\dots,\frac12u)}{1}{(1\frac32)(1\frac12)}
 +
\frac1{\sqrt{6}} \BF{(\frac12r,\dots,\frac12u)}{1}{(1\frac12)(1\frac32)}
 -
\frac{1}{4\sqrt{3}} \BF{(\frac12r,\dots,\frac12u)}{1}{(1\frac32)(1\frac32)}
\\
 & \textstyle
 +
\frac{\sqrt{5}}{4} \BF{(\frac12r,\dots,\frac12u)}{2}{(1\frac32)(1\frac32)}
\,.
\end{align*}
Now, consider the eigenvalue problem for $H$. Expanding
$$
\psi = \sum_J \psi^J \BFi J
 \,,\qquad
{\mathfrak W} = \sum_I W^I (\BFi I + \ol{\BFi I})
 \,,
$$
and using \eqref{StructConst}, as well as the fact that $\braket{\BFi K}{\ol{\BFi I}\BFi J} = \braket{\BFi I \BFi K}{\BFi J} = C_{IK}^J$ implies
$$
\overline{\BFi I} \BFi J = C_{IK}^J \BFi K
\,,
$$
we can write the eigenvalue equation in the form
\beq
\label{EP-H}
\sum_{J \in \mc I} 
 \left\{
\left(\frac{g^2}{2\delta} \epsilon_J - {\cal E} \right) \delta_J^K
 - 
\frac{1}{g^2 \delta} \sum_{I \in \mc I} W^I
\left(C_{IJ}^K + C_{IK}^J\right)\right\} \psi^J
 = 
0
 \,,
\eeq
for all $K \in \mc I$. Here, we have written $\epsilon_J$ for the eigenvalue of the Casimir operator $\mf C$ corresponding to the eigenfunction $\BFi J$, given by \eqref{Eigenvalue-Casimir}. Thus, we are left with a homogeneous system of linear equations for the eigenfunction coefficients $\psi^J$. The eigenvalues $\cal E$ are determined by the requirement that the determinant of this system must vanish. Note that the sum over $I$ in \eqref{EP-H} is finite, because there are only finitely many nonvanishing $W^I$. Moreover, by Remark \ref{finite}, also the sum over $J$ is finite for every fixed $K$. Thus, we have reduced the eigenvalue problem for the Hamiltonian to a problem in linear algebra. Combining this with well-known asymptotic properties of $3nj$ symbols, see \cite{BL2} (Topic 9),\cite{Anderson} and further references therein, we obtain an algebraic setting which allows for a computer algebra supported study of the spectral properties of $H$. This will be done in a future work.


\section{Summary and outlook}

In this paper we have constructed the Hilbert space costratification for $\SU(2)$ lattice gauge theory.  This work is based on the results 
obtained in \cite{FuRS}, where we have implemented the defining relations for the orbit type strata on quantum level.  
Here, the main technical tool is the calculus of invariant representative functions for representations of $\SU(2)$ combined with recoupling 
theory for angular momenta. We have already explained in the introduction how the results of this paper fit into our long-term programme for 
studying non-perturbative aspects of non-abelian quantum gauge theories. Here, let us outline some perspectives:
\begin{enumerate}
\item
It will be a challenge to extend our results to the case of the gauge group $\SU(3)$. On the classical level, we have some preliminary results, see e.g. 
the case studies in \cite{cfg}, \cite{cfgtop} and \cite{FRS}.
\item
In \cite{Fuchs}, one of us has developed another approach towards the study of costratifications for arbitrary compact Lie groups. The starting point in \cite{Fuchs} 
was the observation that the vanishing subspaces corresponding to the classical strata may be viewed as intersections of one-point vanishing subspaces. The orthogonal complements 
of the latter were shown to be one-dimensional and, for each of these spaces, a spanning (holomorphic and square integrable) vector $w_g$ was constructed. Finally, passing to the Hilbert space 
of invariant functions was accomplished by using the projection operator ${\mathbbm P}$ defined by averaging over the compact group manifold.  
As a result, each element of the costratification was characterized as the closure of the span of $\left\{{\mathbbm P w_g}\right\}$, with $g$ running over a complete set of representatives 
of the set of orbits belonging to the stratum under consideration. Moreover, for the point strata, the spanning vectors $w_g$ turned out to be proportional to the 
coherent states in the sense of Hall \cite{Hall:SBT}, \cite{Hall:cptype}. For the other strata, up until now, this approach has not led to such an explicit characterization of 
the corresponding elements of the costratification. It will be interesting to combine the calculus developed in this paper with the methods of \cite{Fuchs}. This will possibly 
lead to a characterization of the full costratification in terms of coherent like states. 
\item
In Subsection \ref{Hamilton}, we have formulated the eigenvalue problem of the quantum Hamiltonian $H$ in terms of invariant representative functions. We have shown that, in this language, 
it boils down to a problem in linear algebra. As already explained there, this can serve as a starting point for a study of the spectral properties of $H$. In particular, 
it should be possible to investigate the role of the coherent states addressed in the previous point, see the toy model in \cite{HRS}.  
\end{enumerate}


\appendix


\begin{section}{Proof of Corollary \ref{F-ReprF}}
\label{Cor}


For every $\lambda \in \widehat G$, we choose a scalar product in $H_\lambda$ invariant under $\pi_\lambda$ and an orthonormal basis $\{e^\lambda_r : r = 1 , \dots , \dim(H_\lambda)\}$. For every $\ul\lambda \in \widehat G^N$, the vectors
$$
e^{\ul\lambda}_{\ul r} := e^{\lambda^1}_{r^1} \otimes \cdots \otimes e^{\lambda^N}_{r^N}
 \,,\qquad
\ul r = (r^1 , \dots , r^N)
 \,,\qquad
r^i= 1 , \dots , \dim(H_{\lambda^i})
\,,
$$
form an orthonormal basis in $H_{\ul\lambda}$ wrt.\ the natural scalar product in the tensor product of Hilbert spaces. Define holomorphic functions 
$$
f^\lambda_{r,s} : G_\CC \to \CC
 \,,\quad
f^\lambda_{r,s}(a)
 := 
\sqrt{\dim(H_\lambda)} \, \braket{e^\lambda_r}{\pi_\lambda(a) e^\lambda_s}
\,,
$$
where $\lambda \in \widehat G$, $r,s = 1 , \dots , \dim(H_\lambda)$, and 
$$
f^{\ul\lambda}_{\ul r,\ul s} : G_\CC^N \to \CC
 \,,\quad
f^{\ul\lambda}_{\ul r,\ul s}(\ul a)
 := 
\sqrt{\dim(H_{\ul\lambda})} \, 
\braket{e^{\ul\lambda}_{\ul r}}{\pi_{\ul\lambda}(\ul a) e^{\ul\lambda}_{\ul s}}
\,,
$$
where $\ul\lambda \in \widehat G^N$ and $\ul r , \ul s \in \prod_{r=1}^N \{1 , \dots , \dim(H_{\lambda^r})\}$. We have 
$$
f^{\ul\lambda}_{\ul r,\ul s}(\ul a)
 =
\prod_{i=1}^N f^{\lambda^i}_{r^i,s^i} (a_i)
\,.
$$
By this relation, the natural unitary isomorphism $\mc HL^2(G^N_\CC) = \big(\mc HL^2(G_\CC)\big)^{\otimes N}$ and the holomorphic Peter-Weyl theorem \cite{HolPW} for $G_\CC$, the functions $f^{\ul\lambda}_{\ul r,\ul s}$ form an orthogonal basis in $\mc HL^2(G^N_\CC)$ and have the norms
$$
\|f^{\ul\lambda}_{\ul r,\ul s}\|^2
 = 
\prod_{i=1}^N \|f^{\lambda^i}_{r^i,s^i}\|^2
 =
\prod_{i=1}^N C_{\lambda^i}\,.
$$
Using the basis vectors $e^{\ul\lambda}_{\ul k}$ in $H_{\ul\lambda}$ to compute the trace, we find the expansion
 \al{\label{G-vaCh-Zlg-C}
\BFC{\ul\lambda}{\lambda}{k,l}
 & =
\sum_{\ul r,\ul s}
\left\langle 
e^{\ul\lambda}_{\ul s}
\right|\left.
A^{\ul\lambda,\lambda}_{k,l} \, e^{\ul\lambda}_{\ul r}
\right\rangle
f^{\ul\lambda}_{\ul r , \ul s}\,.
 }
Using this, orthogonality of the functions $f^{\ul\lambda}_{\ul r , \ul s}$ and orthonormality of the endomorphisms $A^{\ul\lambda,\lambda}_{k,l}$, we obtain
 \ala{
\braket
{\BFC{\ul\lambda}{\lambda}{k,l}}
{\BFC{\ul\lambda'}{\lambda'}{k',l'}}
 & =
\delta_{\ul\lambda , \ul\lambda'} 
\sum_{\ul r , \ul s}
\ol{\left\langle 
e^{\ul\lambda}_{\ul s}
\right|\left.
A^{\ul\lambda,\lambda}_{k,l} \, e^{\ul\lambda}_{\ul r}
\right\rangle}
\left\langle 
e^{\ul\lambda}_{\ul s}
\right|\left.
A^{\ul\lambda,\lambda'}_{k',l'} \, e^{\ul\lambda}_{\ul r}
\right\rangle \, \prod_{i=1}^N C_{\lambda^i}
\\
 & =
\delta_{\ul\lambda , \ul\lambda'} 
 \tr
 \left(
\left(A^{\ul\lambda,\lambda}_{k,l}\right)^\ast 
A^{\ul\lambda,\lambda'}_{k',l'}
 \right)
 \, 
\prod_{i=1}^N C_{\lambda^i}
\\
 & =
\delta_{\ul\lambda , \ul\lambda'} \, \delta_{\lambda\lambda'} \, \delta_{k k'} \, \delta_{l l'} \, \prod_{i=1}^N C_{\lambda^i}
\,.
 }
This yields the assertion.
\qed

\end{section}


\begin{section}{Proof of Lemma \ref{L-symbols} }
\label{Lem}


\bbw

The proof is by induction over $N$. In the case $N=2$, the reduction scheme \eqref{G-Baum-Phi} boils down to \eqref{G-Baum-78} and the reduction scheme \eqref{G-Baum-Psi} boils down to \eqref{G-Baum-36}, where $j_1 = j_1^1$, $j_2 = j_2^1$, $j_3 = j^1$, $j_4 = j_1^2$, $j_5 = j_2^2$, $j_6 = j^2$, $j_7 = l_1^2$, $j_8 = l_2^2$ and $j_9 = l^2$. Hence, by definition, the scalar product of the corresponding ladder basis elements yields the recoupling coefficient
$$
\bpma 
j_1^1 & j_2^1 & j^1 \\ j_1^2 & j_2^2 & j^2 \\ l_1^2 & l_2^2 & l^2
\epma 
$$
and by \eqref{G-U-SU2}, this coincides with $U_{\ul j_1,\ul j_2}(\ul j,\ul l;\ul l_1,\ul l_2)$ in the case at hand. This proves the assertion for $N=2$.

Now, let $N>2$ be given and assume that the assertion holds for $N-1$. In what follows, for any given sequence $\ul x = (x^1 , \dots , x^N)$, we denote $\undertilde x := (x^1 , \dots , x^{N-1})$. Recall that in $H_{\ul j_1} \otimes H_{\ul j_2}$, we have the orthonormal basis vectors $\ket{\ul j_1,\ul j_2;\ul l_1,\ul l_2;l,m}$ defined by the reduction scheme \eqref{G-Baum-Phi} and $\ket{\ul j_1,\ul j_2;\ul j,\ul l,m}$ defined by the reduction scheme \eqref{G-Baum-Psi}. Consider the following reduction scheme: 
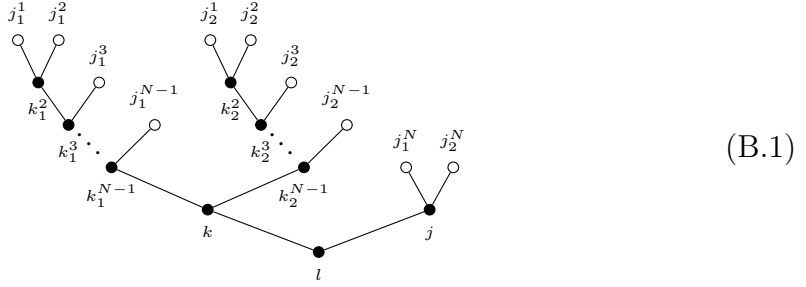
\beq\label{G-L-symbols-1}
\begin{tikzpicture}[baseline = 36pt,grow= up,
  sibling distance =0pt,
  level distance =16pt,
  edge from parent/.style = draw, edge from parent path = {(\tikzparentnode) -- (\tikzchildnode)},
  sloped]
\tikzset{every node/.style = {font = \tiny}},
\tikzset{every internal node/.style ={solid,circle, minimum width = 4pt, fill, inner sep = 0pt, 						draw}},
\tikzset{every leaf node/.style = {solid, circle, minimum width = 4pt, inner sep = 0pt, draw}},
 \Tree
 [.\node[label = below:{$l$}]{};
  [.\node[label = below : {$j$}]{};\node[label = above :{$j_2^N$}]{};\node[label = above :{$j_1^N$}]{};
    ]
  [.\node[label = below : {$k$}]{}; 
  [.\node[label = below :{$k_2^{N-1}$}]{}; \node[label = above :{$j_2^{N-1}$}]{};
    \edge[white] node[black]{\small $\cdots$};
    [.\node[label = below :{$k_2^3$}]{}; \node[label = above :{$j_2^{3}$}]{};
    [.\node[label = below :{$k_2^2$}]{}; \node[label = above:{$j_2^{2}$}]{};\node[label = above:{$j_2^1$}]{};]]]
    [.\node[label = below :{$k_1^{N-1}$}]{}; \node[label = above :{$j_1^{N-1}$}]{};
    \edge[white] node[black]{\small $\cdots$};
    [.\node[label = below :{$k_1^3$}]{}; \node[label = above :{$j_1^{3}$}]{};
    [.\node[label = below :{$k_1^2$}]{}; \node[label = above:{$j_1^{2}$}]{};\node[label = above:{$j_1^1$}]{};]]]]]
\end{tikzpicture}
\eeq
It leads to irreducible subspaces labeled by $\ut k_1 , \ut k_2 , k , j , l$. Let $\ket{\ut k_1 , \ut k_2 ; k , j ; l , m}$ denote the elements of the corresponding orthonormal ladder basis. Using this basis to plug a unit into \eqref{G-U-SU2}, we obtain
 \al{\nonumber
U_{\ul j_1,\ul j_2}(\ul j,\ul l;\ul l_1,\ul l_2;l^N)
 & =
\sum_{\ut k_1,\ut k_2,k,j}
\braket{\ul j_1,\ul j_2;\ul j,\ul l,m}{\ut k_1 , \ut k_2 ; k , j ; l^N , m}
 \times \cdots 
\\ \label{G-L-symbols-2}
 & \hspace{1.5cm}
 \cdots \times
\braket{\ut k_1 , \ut k_2 ; k , j ; l^N , m}{\ul j_1,\ul j_2;\ul l_1,\ul l_2;l^N,m}
\,.
 }
To compute $\braket{\ul j_1,\ul j_2;\ul j,\ul l,m}{\ut k_1 , \ut k_2 ; k , j ; l^N , m}$, we view $H_{\ul j_1} \otimes H_{\ul j_2}$ as the tensor product $\big(H_{\ut j_1} \otimes H_{\ut j_2}\big) \otimes \big(H_{j_1^N} \otimes H_{j_2^N}\big)$ and expand both arguments with respect to appropriately chosen product bases,
 \ala{
\ket{\ul j_1,\ul j_2;\ul j,\ul l,m}
 & =
\sum_{m_1 + m_2 = m}
C^{l^{N-1} j^N l^N}_{m_1 m_2 m}
 \,
\ket{\ut j_1,\ut j_2;\ut j,\ut l,m_1} \otimes \ket{j_1^N,j_2^N;j^N,m_2}
\\
\ket{\ut k_1 , \ut k_2 ; k , j ; l^N , m}
 & =
\sum_{m_1 + m_2 = m}
C^{k\ j\ l^N}_{m_1 m_2 m}
 \,
\ket{\ut j_1,\ut j_2;\ut k_1,\ut k_2;k,m_1} \otimes \ket{j_1^N,j_2^N;j,m_2}
\,.
 }
In view of \eqref{G-U-SU2}, this yields
\beq\label{G-L-symbols-3}
\braket{\ul j_1,\ul j_2;\ul j,\ul l,m}{\ut k_1 , \ut k_2 ; k , j ; l^N , m}
 =
\delta_{l^{N-1},k} \, \delta_{j^N,j}
 \,\,
U_{\ut j_1,\ut j_2}\left(\ut j,\ut l;\ut k_1,\ut k_2,k\right)
\,.
\eeq
To compute the scalar product $\braket{\ut k_1 , \ut k_2 ; k , j ; l^N , m}{\ul j_1,\ul j_2;\ul l_1,\ul l_2;l^N,m}$, we observe that the vectors $\ket{\ut k_1 , \ut k_2 ; k , j ; l^N , m}$ are ladder basis elements in the tensor product\footnote{See \eqref{G-D-Hjl} for the notation $H_{\ut j_1,\ut k_1}$ etc.\ .} 
$$
H_{\ut j_1,\ut k_1} \otimes H_{\ut j_2,\ut k_2} \otimes H_{j_1^N} \otimes H_{j_2^N}
$$
defined by the reduction scheme \eqref{G-Baum-36} with $j_1 = k_1^{N-1}$, $j_2 = k_2^{N-1}$, $j_3 = k$, $j_4 = j_1^N$, $j_5 = j_2^N$, $j_6 = j$ and $j_9 = l^N$, whereas $\ket{\ul j_1,\ul j_2;\ul l_1,\ul l_2;l^N,m}$ are ladder basis elements in the tensor product 
$$
H_{\ut j_1,\ut l_1} \otimes H_{\ut j_2,\ut l_2} \otimes H_{j_1^N} \otimes H_{j_2^N}
$$
defined by the reduction scheme \eqref{G-Baum-78} with $j_1 = l_1^{N-1}$, $j_2 = l_2^{N-1}$, $j_4 = j_1^N$, $j_5 = j_2^N$, $j_7 = l_1^N$, $j_8 = l_2^N$ and $j_9 = l^N$. Therefore,
$$
\braket{\ut k_1 , \ut k_2 ; k , j ; l^N , m}{\ul j_1,\ul j_2;\ul l_1,\ul l_2;l^N,m}
 =
\delta_{\ut k_1 , \ut l_1} \, \delta_{\ut k_2 , \ut l_2} \, 
\bpma l_1^{N-1} & l_2^{N-1} & k \\ j_1^N & j_2^N & j \\ l_1^N & l_2^N & l^N \epma
\,.
$$
Plugging this and \eqref{G-L-symbols-3} into \eqref{G-L-symbols-2} and taking the sum, we obtain
$$
U_{\ul j_1,\ul j_2}(\ul j,\ul l;\ul l_1,\ul l_2)
 =
\bpma l_1^{N-1} & l_2^{N-1} & l^{N-1} \\ j_1^N & j_2^N & j^N \\ l_1^N & l_2^N & l^N \epma
U_{\ut j_1,\ut j_2}\left(\ut j,\ut l;\ut l_1,\ut l_2\right)
\,.
$$
Thus, the induction assumption implies that the assertion holds for $N$.
\ebw

\end{section}


\end{document}